\documentclass{aa}  

\usepackage{graphicx}
\usepackage{txfonts}
\usepackage{aalongtable}
\usepackage{amsmath}
\begin{document}

   \title{KIC\,10080943: An eccentric binary system containing two pressure- and gravity-mode hybrid pulsators\thanks{Based on the data gathered with NASA's Discovery mission, \textit{Kepler,} and with the HERMES spectrograph, installed at the Mercator Telescope, operated on the island of La Palma by the Flemish Community, at the Spanish Observatorio del Roque de los Muchachos of the Instituto de Astrof\'{i}sica de Canarias and supported by the Fund for Scientific Research of Flanders (FWO), Belgium, the Research Council of KU\,Leuven, Belgium, the Fonds National de la Recherche Scientific (F.R.S.--FNRS), Belgium, the Royal Observatory of Belgium, the Observatoire de Gen\`{e}ve, Switzerland, and the Th\"{u}ringer Landessternwarte Tautenburg, Germany.}}
   \titlerunning{KIC\,10080943: A binary containing two hybrid pulsators.}

   \author{V.\,S.~Schmid\inst{1}\fnmsep\thanks{Aspirant PhD Fellow of the Fund for Scientific Research of Flanders (FWO), Belgium}
          \and
          A.~Tkachenko\inst{1}\fnmsep\thanks{Postdoctoral Fellow of the Fund for Scientific Research of Flanders (FWO), Belgium}
          \and
          C.~Aerts\inst{1,2}
          \and
          P.~Degroote\inst{1}\fnmsep$^{\star\star\star}$
          \and
          S.~Bloemen\inst{2}
          \and
          S.\,J.~Murphy\inst{3,4}
          \and
          T.~Van~Reeth\inst{1}
          \and
          P.\,I.~P\'{a}pics\inst{1}\fnmsep$^{\star\star\star}$
          \and
          T.\,R.~Bedding\inst{3,4}
          \and
          M.\,A.~Keen\inst{3}
          \and
          A.~Pr\v{s}a\inst{5}
          \and
          J.~Menu\inst{1}\fnmsep$^{\star\star}$
          \and
          J.~Debosscher\inst{1}
          \and
          M.~Hrudkov\'{a}\inst{6}
          \and
          K.~De~Smedt\inst{1}
          \and
          R.~Lombaert\inst{1}
          \and
          P.~N\'{e}meth\inst{7}
          }

   \institute{
   Institute for Astronomy, KU Leuven, Celestijnenlaan 200D, B -- 3001 Leuven, Belgium\\
   \email{valentina.schmid@ster.kuleuven.be}
   \and
   Department of Astrophysics/IMAPP, Radboud University Nijmegen, P.O. Box 9010, 6500 GL Nijmegen, The Netherlands
   \and
   Sydney Institute for Astronomy (SIfA), School of Physics, The University of Sydney, NSW 2006 Australia
   \and
   Stellar Astrophysics Centre, Department of Physics and Astronomy, Aarhus University, 8000 Aarhus C, Denmark
   \and
   Department of Astronomy and Astrophysics, Villanova University, 800 E. Lancaster Avenue, Villanova, PA 19085, USA
   \and
   Isaac Newton Group of Telescopes, Apartado de Correos 321, E-38700 Santa Cruz de la Palma, Canary Islands, Spain
   \and
   Dr. Karl Remeis-Observatory \& ECAP, Astronomisches Inst., FAU Erlangen-Nuremberg, 96049 Bamberg, Germany
   }

   \date{Received 13 July 2015 / Accepted 23 August 2015 }

 
  \abstract  
  { Gamma Doradus and delta Scuti pulsators cover the transition region between low mass and massive main-sequence stars, and as such, are critical for testing stellar models. When they reside in binary systems, we can combine two independent methods to derive critical information, such as precise fundamental parameters to aid asteroseismic modelling. In the \textit{Kepler} light curve of KIC\,10080943, clear signatures of gravity- and pressure-mode pulsations have been found. Ground-based spectroscopy revealed this target to be a double-lined binary system. }
   {  We present the analysis of four years of \textit{Kepler} photometry and high-resolution spectroscopy to derive observational constraints with which to evaluate theoretical predictions of the stellar structure and evolution for intermediate-mass stars. }
   { We used the method of spectral disentangling to determine atmospheric parameters for both components and derive the orbital elements. With \textsc{phoebe,} we modelled the ellipsoidal variation and reflection signal of the binary in the light curve and used classical Fourier techniques to analyse the pulsation modes. }
   {We show that the eccentric binary system KIC\,10080943 contains two hybrid pulsators with masses $M_1=2.0\pm0.1~M_\sun$ and $M_2=1.9\pm0.1~M_\sun$, with radii $R_1=2.9\pm0.1~R_\sun$ and $R_2=2.1\pm0.2~R_\sun$. We detect rotational splitting in the g and p~modes for both stars and use them to determine a first rough estimate of the core-to-surface rotation rates for the two components, which will be improved by future detailed seismic modelling.  }
   {}


   \maketitle
%

\section{Introduction}

The stellar interior was one of the most inaccessible parts of the universe until the advent of asteroseismology. Nowadays we can use stellar pulsations to inspect the conditions deep beneath the stellar photosphere and to calibrate theoretical stellar models \citep[e.g.][]{Chaplin2013,Aerts2015}. A particularly interesting group of stars for this type of analysis are the $\gamma$\,Doradus (hereafter $\gamma$\,Dor) and $\delta$\,Scuti (hereafter $\delta$\,Sct) pulsators. They are found on and slightly above the main sequence, near the classical instability strip of the Hertzsprung-Russell diagram (HRD), ranging between $1.5~M_\sun$ and $2.5~M_\sun$ in mass. In this transition region, the outer convective envelope becomes shallower with increasing mass until it is negligible, and energy is mainly transported radiatively, while in the centre, the core becomes increasingly convective.

The $\gamma$\,Dor stars are somewhat cooler and less massive than $\delta$\,Sct stars, having masses from $1.5~M_\sun$ to $1.8~M_\sun$ and spectral type A7-F5 \citep[and references therein]{Kaye1999}. Their typical pulsation periods between 0.3 d and 3 d are consistent with non-radial, high-order, low-degree gravity (g) modes, which are driven by the convective flux blocking mechanism at the base of the convective envelope \citep{Guzik2000, Dupret2005}. The $\delta$\,Sct stars, on the other hand, have been studied for almost a century and have masses up to $2.5~M_\sun$. They pulsate in radial as well as non-radial, low-order, low-degree pressure (p) modes at shorter periods between 18 min and 8 hr, which are driven by the $\kappa$ mechanism \citep[e.g.][]{Breger2000a}. Besides main-sequence stars and more evolved sub-giants, $\delta$\,Sct pulsations have also been detected in pre-main sequence stars, providing an important tracer for early stellar evolution \citep{Zwintz2014}. Where the $\gamma$\,Dor and $\delta$\,Sct instability strips overlap, stars are expected to show both types of pulsations \citep{Dupret2005}. These so-called hybrid pulsators are especially useful for asteroseismic studies to test stellar models, since the g-mode cavity is located close to the core and p~modes have higher amplitudes in the outer envelope.

For a non-rotating star, it is theoretically predicted that high-degree g~modes of the same degree $\ell$ and of consecutive radial orders $n$ are equidistantly spaced in period if $n\gg\ell$ \citep{Tassoul1980}. \citet{Miglio2008} demonstrate that these period spacings can depart from their equidistant behaviour as the star evolves from the start to the end of its core hydrogen-burning phase. These departures from regularity carry crucial information on the conditions near the core where g~modes have their largest amplitudes. A gradient in chemical composition can cause mode trapping and can lead to an oscillatory behaviour of period spacings, while additional mixing processes, such as diffusion, convective-core overshooting, and rotation can reduce this gradient. Additional mixing mechanisms can significantly influence the time a star spends on the main sequence by supplying the core with additional hydrogen for fusion \citep{Miglio2008}. 

Rotation can furthermore introduce rotational splitting, which facilitates identification of the modes' azimuthal order $m$ and can shift g~modes to shorter periods, which has been studied extensively by \citet{Bouabid2013}. Rotational splitting, as first described by \citet{Ledoux1951}, also occurs for p~modes. If it is detected in hybrid pulsators in both the g and p~modes, additional constraints can be derived on the radial rotation profile in a largely model-independent way. This has recently been achieved by \citet{Kurtz2014} and \citet{Saio2015}, who found nearly rigid rotation in two hybrid pulsators, requiring a much more efficient angular momentum transport mechanism than previously assumed. This has also been concluded from surface-to-core rotation rates detected for red giant stars through rotationally split mixed modes \citep{Beck2012,Mosser2012,Deheuvels2014,Deheuvels2015}, which are two orders of magnitudes below the theoretically predicted value \citep{Cantiello2014,Fuller2015}. For the Sun, on the other hand, it has not yet been possible to measure the rotation of the core, as until now, no g-mode pulsations have been observed .

These and other recent breakthroughs in asteroseismology have been achieved by exploiting the nearly uninterrupted, high precision photometry of space missions, such as MOST \citep{Walker2003}, CoRoT \citep{Auvergne2009}, and \textit{Kepler} \citep{Borucki2010}. In particular, the detection and analysis of $\gamma$\,Dor stars and hybrid pulsators have benefited from these observations, since they have been studied in extensive samples of A-F type stars observed with \textit{Kepler} \citep[e.g.][]{Grigahcene2010,Uytterhoeven2011,Balona2014,Bradley2015}. In addition, \citet{Tkachenko2013a} selected 69 candidate $\gamma$\,Dor stars, which have been followed up spectroscopically and analysed in more detail by \citet{VanReeth2015a,VanReeth2015b}. Facilitated by the high frequency resolution possible from four years of \textit{Kepler} photometry, they found period spacing patterns in 50 stars. Four more g-mode pulsators have been presented by \citet{Bedding2014}, who detect period spacings and rotational multiplets from period \'{e}chelle diagrams. These excellent observational results confirm the theoretical predictions of period dips, which are the result of a chemical gradient, and slopes, which are the result of rotation, by \citet{Bouabid2013}.

Despite the immense potential of asteroseismic modelling for stellar astrophysics, a unique solution is hampered by degeneracies among the free parameters \citep[e.g.\ mass, age, initial metallicity, chemical mixture, and mixing processes; see, e.g.][]{Moravveji2015}. Additional observational constraints to help lift these degeneracies can be provided by binary stars that contain at least one pulsating component. In particular, eclipsing binaries provide the opportunity to derive masses and radii to very high precision (up to $1\%$), as they rely purely on geometry \citep{Southworth2012} and, therefore, aid conclusions derived from modelling. Furthermore, eccentric binaries can tidally excite eigenmodes or alter the frequencies or amplitudes of free pulsations \citep{Maceroni2009,Welsh2011,Hambleton2013}, which provide additional opportunities for stellar modelling \citep{Fuller2012}. In addition, \citet{Welsh2011} were able to determine the orbital inclination, and therefore the absolute masses of the non-eclipsing binary KOI\,54 by modelling ellipsoidal variation (the deformation of stellar surface models as Roche potentials) and reflection \citep[mutual heating of irradiated stellar surfaces;][]{Wilson1990}. However, despite their obvious benefits, pulsating binaries also present a challenging analysis, in which signals of two different origins must be disentangled. A common approach has been to treat the asteroseismic and binary signals separately to solve the system iteratively, which has been applied, for example, by \citet{Maceroni2013,Maceroni2014} and \citet{Debosscher2013}.

In this paper, we analyse the eccentric, non-eclipsing binary star KIC\,10080943 \citep[effective temperature $T_\mathrm{eff} \sim 7400$~K; surface gravity $\log g\sim 4.0$;][]{Huber2014}, observed during all four years of the nominal \textit{Kepler} mission. \citet{Tkachenko2013a} discovered pulsations of $\gamma$\,Dor and $\delta$\,Sct type, indicating that the system might contain a hybrid pulsator. From ground-based follow-up observations, they further discovered that it is a double-lined spectroscopic binary.

In Sect.~\ref{sec:observations} we describe the acquisition, characteristics, and reduction of the \textit{Kepler} photometry and ground-based spectroscopy, which we use to derive atmospheric parameters in Sect.~\ref{sec:spectroscopy}. We describe the modelling of the binary light curve (Sect.~\ref{sec:binary}), as well as the frequency analysis (Sect.~\ref{sec:freqanalysis}), before discussing our findings in Sect.~\ref{sec:discussion} and ending with a brief summary. The analysis of the g modes is presented in an earlier, separate paper \citep{Keen2015}.


\section{Observations}
\label{sec:observations}
\subsection{\textit{Kepler} photometry}

\begin{figure*}
\centering
\includegraphics{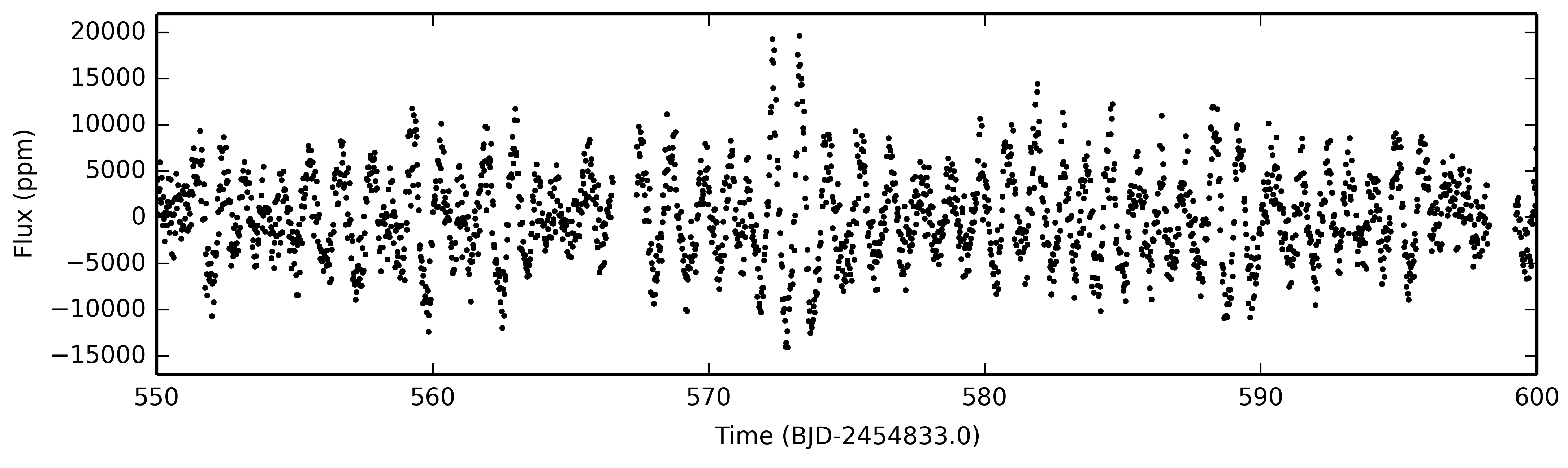}
\caption{Zoom of 50 days of the \textit{Kepler} light curve of KIC\,10080943.}
\label{fig:lightcurve}
\end{figure*}

The \textit{Kepler} satellite was launched on 6 March 2009 and ended its nominal mission on 11 May 2013, after a second reaction wheel failed and pointing it to its chosen field could no longer be kept stable. During the course of its operation, \textit{Kepler} collected data in eighteen quarters, employing a sampling rate of 29.4244 min in long cadence (LC) mode. For KIC10080943, there are LC observations available from Q0 to Q17, amounting to a time span of 1470.5 days, which contain 65\,959 data points after deleting 22 outliers. We extracted the light curve from the public pixel data files using a custom aperture mask, which contained more pixels than the original mask to minimise long term trends \citep{Tkachenko2013a}. After each quarter, the satellite rolled to adjust the position of the solar panels. This caused short gaps in the light curve and made the stars fall on different CCD modules, which deviate in sensitivity and lead to varying observed flux. By fitting a second order polynomial to the light curve in each quarter, we rectify any spurious trends to safely concatenate the data \citep[for more details see][]{Debosscher2013}. A close-up of the final light curve is shown in Fig.~\ref{fig:lightcurve}, where time is given as Barycentric \textit{Kepler} Julian Date (BKJD), which is Barycentric Julian Date (BJD), using a zero point of 2454833.0.

\subsection{HERMES spectroscopy}

In addition to the photometric observations, we monitored the star with the HERMES spectrograph \citep{Raskin2011}, mounted on the 1.2-m Mercator telescope in La Palma, Spain. We obtained 26 spectra between August 2011 and October 2014. The raw spectra were reduced using the HERMES pipeline. Since the target is faint for a high-resolution spectrograph on a 1.2-m telescope (\textit{Kepler} magnitude $Kp = 11.8$~mag), most spectra that were obtained have a signal-to-noise ratio ($S/N$) below $30$. We prepared the spectra for normalisation by removing the spectrograph's response function and then dividing this by a synthetic composite spectrum of two identical stars, given the component's similarity and their mass ratio close to unity ($T_\mathrm{eff,1,2}=7000$~K, $\log g_\mathrm{1,2} = 4.0$, metallicity $Z_\mathrm{1,2}=0$, projected rotational velocity of the primary $v\sin i_\mathrm{1}=13$~km\,s$^{-1}$, and of the secondary $v\sin i_\mathrm{2}=10$~km\,s$^{-1}$). Subsequently, we fitted the result with a third-order polynomial, avoiding regions of hydrogen and telluric lines, and used it to normalise the spectra. Given that double-lined F-type binary spectra contain a large number of lines and the observations have a low $S/N$, we found this method to be the most reliable. Our tests showed that fitting a second-order spline to continuum points resulted in a wavy continuum, varying between spectra. 

We used the normalised spectra to perform spectral disentangling and to obtain fundamental parameters for both components (see Sect.~\ref{sec:spectroscopy}). In addition, we extracted the radial velocities (RVs) of both components from the spectra's \'{e}chelle orders prior to normalisation by cross-correlating with a mask of an F0 star and fitting a double Gaussian to the resulting function.

\section{Spectral disentangling, binary orbit and atmospheric parameters}
\label{sec:spectroscopy}

\begin{figure}
\centering
\includegraphics{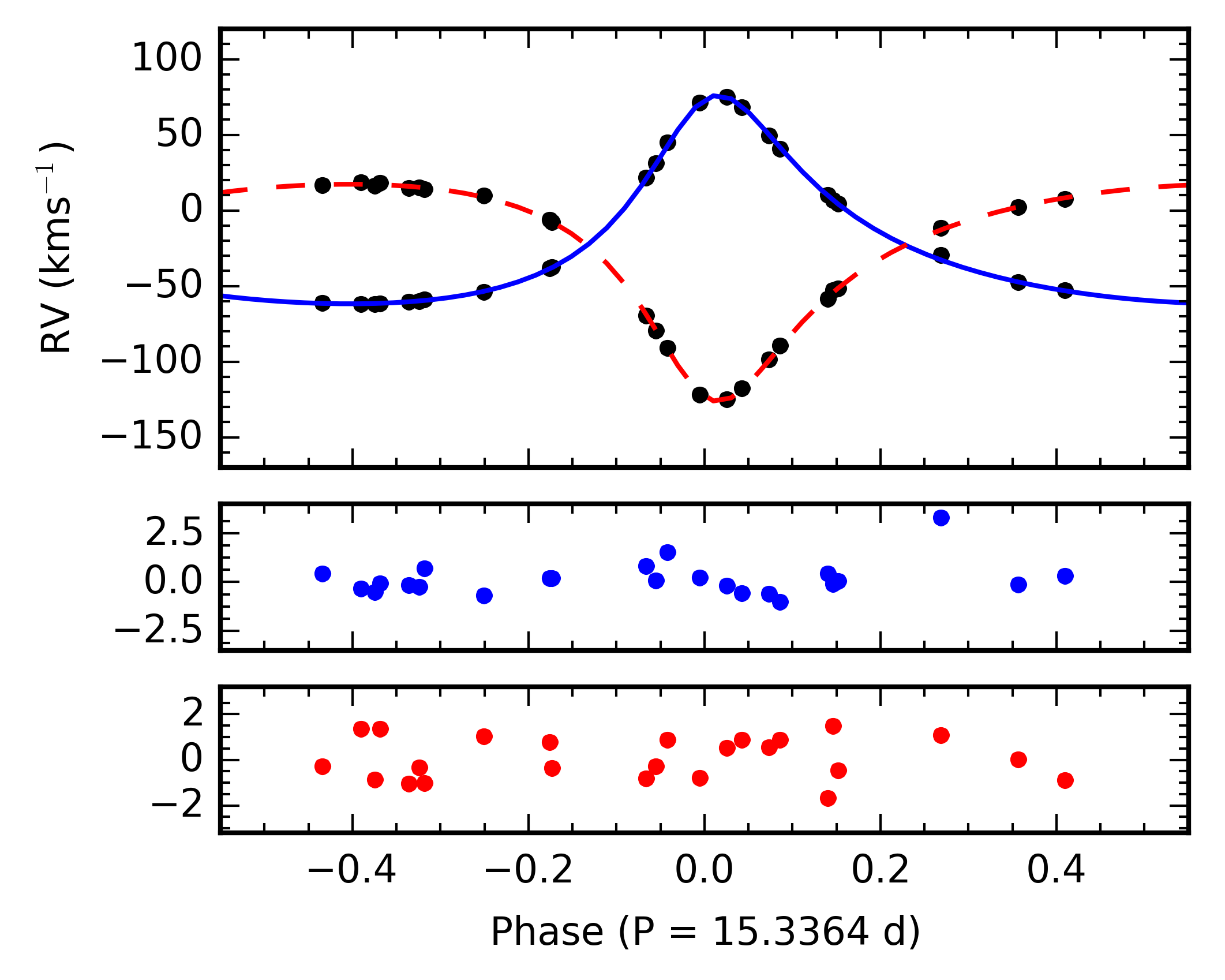}
\caption{\textit{Top panel:} Radial velocities (black dots) of both components with the best fitting model for the primary (solid blue line) and the secondary (dashed red line). The residuals after subtracting the best fit are displayed for the primary in the middle panel and the secondary in the bottom panel. The error bars on each RV measurement are typically smaller than the symbol size.}
\label{fig:RV_fit}
\end{figure}

We used the Fourier method of spectral disentangling \citep{Hadrava1995}, as implemented in the FDBinary code \citep{Ilijic2004}, to extract the spectral contributions of both binary components from the observed composite spectra of the system. The method \citep[originally introduced by][]{Simon1994} allows for simultaneous separation of the spectra of the two binary components and optimisation of the system's orbital elements. Before determining the orbit, we used the RVs to derive the orbital period, which was then fixed in FDBinary to simplify the optimisation problem. The following five spectral regions were used to calculate the orbital parameters: 4530 -- 4590~$\mathrm{\AA}$, 4600 -- 4650~$\mathrm{\AA}$, 5164 -- 5174~$\mathrm{\AA}$, 5250 -- 5295~$\mathrm{\AA}$, and 5350 -- 5380~$\mathrm{\AA}$. Table~\ref{tab:binary} lists the mean and standard deviation of the parameters: time of periastron passage $t_0$, eccentricity $e$, longitude of periastron $\omega$, and the RV semi-amplitudes $K_1$ and $K_2$. The mass ratio $q=M_1/M_2=K_2/K_1$, the systemic velocity $\gamma$, and the semi-major axis as a function of orbital inclination $i_\mathrm{orb}$, $a\sin i_\mathrm{orb}$, are also listed. All values are in agreement with those derived from fitting the RVs. The orbital fit to the RVs is shown in Fig.~\ref{fig:RV_fit}.

In the next step, we fixed the orbital elements to the mean values and performed spectral disentangling in the range between 4200~$\mathrm{\AA}$ and 5650~\AA. The violet part of the spectrum was ignored because of large uncertainties in the continuum normalisation, while the red part was skipped because of large telluric contamination and an increasing level of noise. The quality of the continuum normalisation of the original data suffered significantly from the low $S/N$ of the data. This naturally propagated into the quality of the disentangled spectra, particularly in the regions of the Balmer lines. We found that the normalisation of the H$_{\beta}$ profile was not accurate enough to provide reliable Fourier spectral disentangling in the corresponding wavelength region. Our attempt to perform disentangling in the wavelength domain, using the original method of \citet{Simon1994}, also failed. The disentangling went reasonably well for the entire red part of the H$_{\gamma}$ profile, whereas its blue wing suffered significantly from the zero-frequency component in the Fourier domain. The blue wing was corrected by fitting a low-degree polynomial, to restore the symmetry of the entire profile with respect to its red wing. The reliability of such an approach was later tested by means of the spectrum analysis based on the entire wavelength range and the one excluding the H$_{\gamma}$ profile. The metal line spectrum was disentangled in small segments ($\sim$30 -- 50~\AA\ each), where low-amplitude continuum undulations were corrected by fitting a low-degree polynomial to each of the wavelength regions. The final disentangled spectra were obtained by merging all segments.

\begin{figure*}
\includegraphics[scale=1.2]{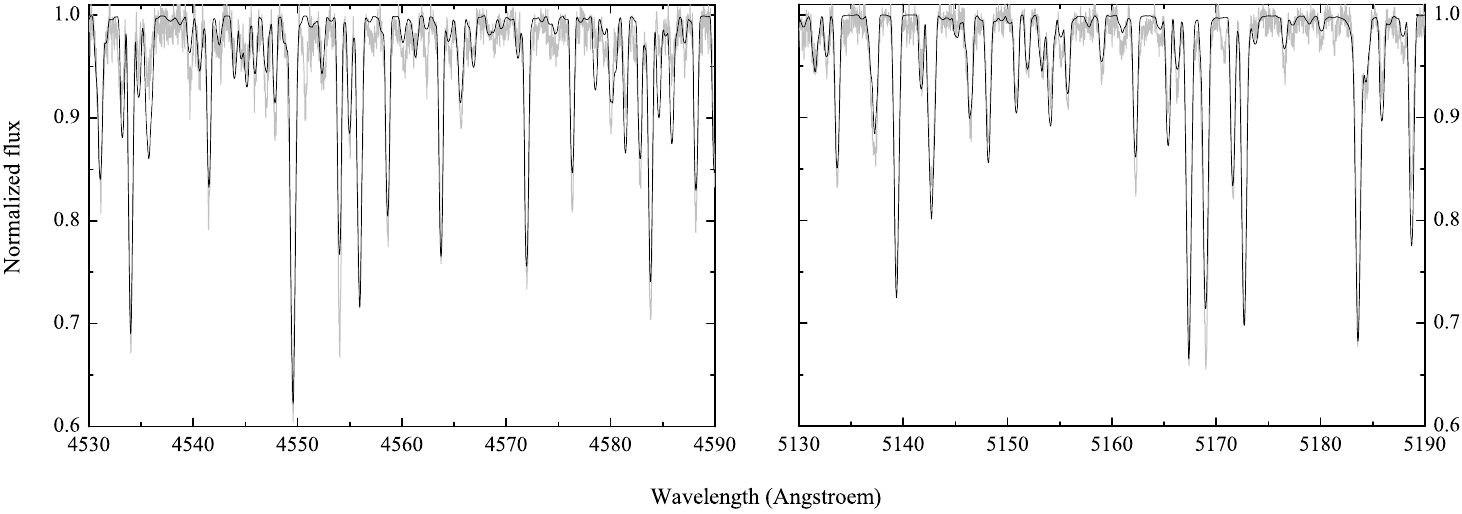}
\includegraphics[scale=1.2]{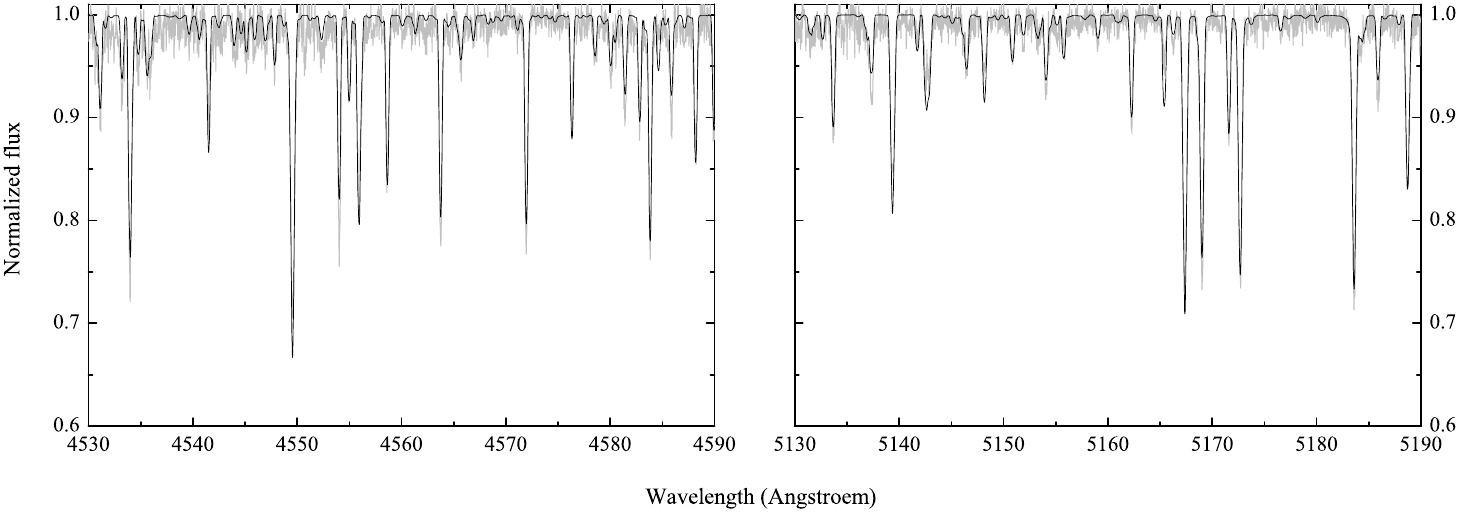}
\caption{Fit to the disentangled spectra of both components of the KIC\,10080943 system (top panels - primary, bottom
panels - secondary) with the \textsc{gssp\_binary} software package. The observations are shown as the light grey line and the best fit synthetic spectra are illustrated with the black line.}
\label{fig:spectralanalysis}
\end{figure*}

We used the \textsc{gssp\_binary} software package \citep{Tkachenko2015} to fit the disentangled spectra. The spectrum analysis relies on a grid search in the fundamental parameters: $T_\mathrm{eff}$, $\log g$, micro-turbulence $\xi$, metallicity [M/H], projected rotational velocities $v\sin i_{1,2}$, and the ratio of the radii $R_1/R_2$; the 1-$\sigma$ uncertainties are calculated from $\chi^2$ statistics, taking into account all possible correlations between parameters \citep{Lehmann2011}. Due to the low $S/N$ of the data, micro-turbulence is hard to constrain and we derived another solution with micro-turbulence fixed at $2.0~\mathrm{km\,s}^{-1}$. The code allows for constrained fitting of the two spectra simultaneously and takes into account the wavelength dependence of the light ratio of the two stars, by replacing it with the ratio of their radii. We found that the final solution was very sensitive to whether the H$_\gamma$ spectral line was included in the fit, particularly influencing $T_\mathrm{eff}$ and $\log g$ of the primary component. Given the level of uncertainty in the disentangling of this line and the corrections applied to its blue wing afterwards, we decided to exclude it from the final fit and focused on the metal line spectrum for each of the components. This approach is justified by the large number of metal lines present in the spectra of both stars, which contain sufficient information for $T_\mathrm{eff}$ and $\log g$ to be constrained from the excitation and ionisation balance, respectively. The finally adopted fundamental parameters are listed in Table~\ref{tab:binary}. The quality of the fit for both stars in two different wavelength regions is illustrated in Fig~\ref{fig:spectralanalysis}. The more-massive primary component is found to have a $v\sin i$ about 1.35 times greater than the secondary, which is in turn slightly hotter. A less-massive but hotter secondary is counter-intuitive, but can be explained by a larger primary radius. We find that the components' metallicities are consistent with each other and may be slightly sub-solar \citep[the solar composition was adopted from][]{Grevesse2007}.

\begin{table}
\caption{Parameters of the binary orbit and the component stars of KIC\,10080943.}
\label{tab:binary}
\centering
\def\arraystretch{1.3}
\begin{tabular}{l c c}
\hline\hline 
 & \textbf{Primary} & \textbf{Secondary} \\ 
\hline
$P_\mathrm{orb}$ (d) & \multicolumn{2}{c}{$15.3364\pm0.0003$} \\
$f_\mathrm{orb}$ (d$^{-1}$) & \multicolumn{2}{c}{$0.06520\pm0.00004$} \\
\hline
\multicolumn{3}{l}{\textbf{FDBinary}} \\
\hline
$t_0$ (JD) & \multicolumn{2}{c}{$2\,455\,782.23\pm0.02$} \\
$e$ & \multicolumn{2}{c}{$0.449\pm0.005$} \\
$\omega$ (deg) & \multicolumn{2}{c}{$344.7\pm0.7$} \\
$a\sin i_{\mathrm{orb}}$ ($R_\sun$) & \multicolumn{2}{c}{$37.8\pm0.3$} \\
$q$ & \multicolumn{2}{c}{$0.96\pm0.01$} \\
$K$ (km\,s$^{-1}$) & $68.2\pm0.7$ & $71.3\pm0.7$ \\
$\gamma$ (km\,s$^{-1}$) & \multicolumn{2}{c}{$-22.8\pm0.2$} \\
\hline
\multicolumn{3}{l}{\textbf{Phoebe}} \\
\hline
$i_{\mathrm{orb}}$ (deg) & \multicolumn{2}{c}{$68\pm3$} \\
$\Delta \phi$ & \multicolumn{2}{c}{$0.2928\pm0.0001$} \\
$e$ & \multicolumn{2}{c}{$0.4535\pm0.0004$} \\
$\omega$ (deg) & \multicolumn{2}{c}{$345.309\pm0.001$} \\
$a$ ($R_\sun$) & \multicolumn{2}{c}{$41.1\pm0.8$} \\
$q$ & \multicolumn{2}{c}{$0.9598^{+0.0007}_{-0.0008}$} \\ 
$\gamma$ (km\,s$^{-1}$) & \multicolumn{2}{c}{$-22.99\pm0.02$} \\
$T_\mathrm{eff}$ (K) & $7100\pm200$ & $7480^{+180}_{-200}$ \\
$\Omega$ & $15.2\pm0.2$ & $23.6^{+0.9}_{-1.0}$ \\
$\alpha$ & $0.8\pm0.1$ & $0.95\pm0.05$ \\
$\beta$ & $0.04^{+0.06}_{-0.03}$ & $0.27^{+0.34}_{-0.21}$ \\
$M$ ($M_\sun$) & $2.0\pm0.1$ & $1.9\pm0.1$ \\
$R$ ($R_\sun$) & $2.9\pm0.1$ & $2.1\pm0.2$ \\
$\log g$ (cgs) & $3.81\pm0.03$ & $4.1\pm0.1$ \\
$R_1/R_2$ & \multicolumn{2}{c}{$1.4\pm0.1$} \\
\hline
\multicolumn{3}{l}{\textbf{GSSP} (excluding $H_\gamma$)} \\
\hline
$T_\mathrm{eff}$ (K) & $7150\pm250$ & $7640\pm240$ \\
$\log g$ (cgs) & $4.07\pm0.46$ & $4.13\pm0.46$ \\
$\xi$ (km\,s$^{-1}$) & $3.85\pm0.90$ & $2.35\pm0.85$ \\
$\mathrm{[M/H]}$ & $-0.06\pm0.18$ & $-0.23\pm0.20$ \\
$v\sin i_{\mathrm{rot}}$ (km\,s$^{-1}$) & $19.0\pm1.3$ & $14.4\pm1.4$ \\
$R_1/R_2$ & \multicolumn{2}{c}{$1.16\pm0.19$} \\
\multicolumn{3}{c}{\textit{Micro-turbulence fixed at 2.0 km\,s$^{-1}$}} \\
$T_\mathrm{eff}$ (K) & $7100\pm220$ & $7450\pm400$ \\
$\log g$ (cgs) & $3.7\pm0.4$ & $4.15\pm0.85$ \\
$\mathrm{[M/H]}$ & $-0.05\pm0.17$ & $-0.09\pm0.30$ \\
$v\sin i_{\mathrm{rot}}$ (km\,s$^{-1}$) & $18.7\pm1.2$ & $13.8\pm1.6$ \\
$R_1/R_2$ & \multicolumn{2}{c}{$1.41\pm0.20$} \\
\hline
\end{tabular}
\end{table}

\section{Binary light curve modelling}
\label{sec:binary}

\begin{figure}
\centering
\includegraphics{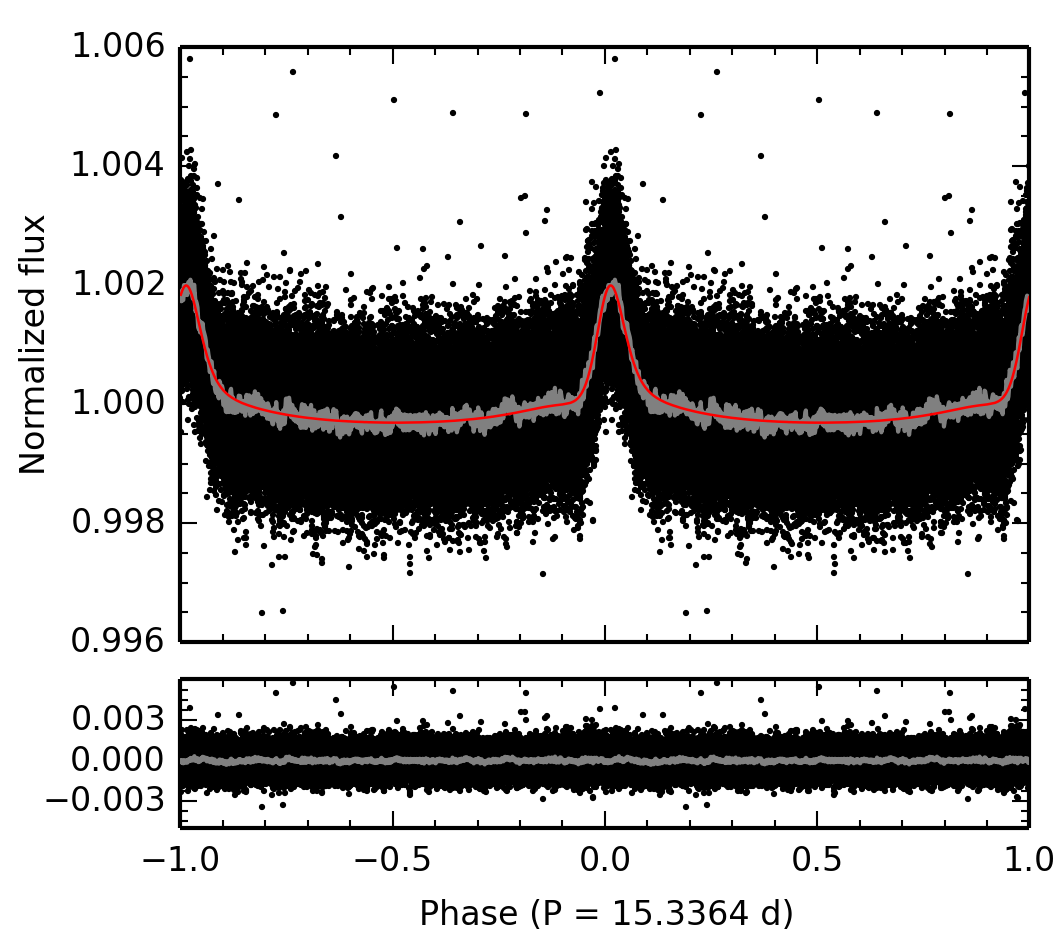}
\caption{Phased binary signal in the observed light curve with oscillations removed (black), and the best fitting model (red) with the residuals shown below. The 1-$\sigma$ area of the phase-binned light curve is shown in grey.}
\label{fig:lc_fit}
\end{figure}

The binary signal also appears in the \textit{Kepler} light curve. The orbit is seen in the Fourier transform as the first ten harmonics of the orbital frequency, $f_\mathrm{orb}=0.0652~\mathrm{d}^{-1}$, with amplitudes well below most pulsation modes (Fig.~\ref{fig:pergram}; panel b). Ellipsoidal variation and reflection are non-sinusoidal signals at high eccentricity and cause harmonic peaks at integer multiples of the orbital frequency. Subtracting the g- and p-mode pulsations using classical iterative prewhitening (see Sect.~\ref{sec:freqanalysis}) from the original light curve reveals a periodic brightening occurring at orbital phase zero, which we associate with ellipsoidal variation and reflection (Fig.~\ref{fig:lc_fit}). To complete the iterative approach of studying pulsating binary light curves, we modelled this signal and subsequently subtracted it from the original light curve, to extract the pure pulsation signal. This procedure was only repeated once, as the difference in subsequent frequency and binary fits was negligible.

\subsection{MCMC setup}

Since KIC\,10080943 does not show eclipses, the modelling is not well constrained and degeneracies occur among parameters, such as effective temperatures and stellar surface potentials, $\Omega$, which define the stars' shapes in Roche geometry and can be used to derive the stellar radii. The shape of the ellipsoidal variation signal is, in principle, a function of the temperature ratio, stellar radii, eccentricity, and inclination. At higher orbital inclination, we expect to see a double-humped feature, while for low values of inclination, only one hump appears. Before modelling, we concluded that $i_\mathrm{orb}$ probably lies between $60^{\degr}$ and $82.5^{\degr}$, since a lower value would yield masses that are too high, given an early-F spectral type, and a higher value would result in eclipses. The fact that we only see one peak in the light curve means that reflection is the dominating effect, which constrains the sizes of the stars and their temperatures. Furthermore, this brightening signal has a small amplitude and, thus, a low $S/N$. Small variations in the light curve will therefore be drowned out in the noise and remaining unresolved pulsation frequencies. In this case, the parameter hypersurface will have many dips and troughs, and minimisation algorithms might get stuck in local minima.

To overcome these obstacles we perform a Markov Chain Monte Carlo (MCMC) simulation \citep[implemented in \texttt{emcee} by][]{Foreman-Mackey2013}, which allows several chains of models to explore the parameter space and build up a posterior probability density function of the parameter space $\Theta$, given the data set $D$. If we assume that our parameters follow a Gaussian distribution, we can write the posterior probability as
\begin{equation}
P(\Theta | D) = P(\Theta)e^{-\chi^2/2},
\end{equation}
where $\chi^2$ is a simple goodness-of-fit measurement. The $\chi^2$ values were calculated at each step of each chain using binary models computed with \textsc{phoebe} \citep{Prsa2005}, which is based on the Wilson-Divinney code \citep{Wilson1971,Wilson1979}. In this way, an MCMC simulation also improves the model fit to the data by maximising the posterior. Gaussian priors of the effective temperatures and the orbital parameters ($e$, $\omega$, $q$, and $\gamma$), which are based on the spectral analysis and spectral disentangling, respectively, were used to write the posterior probability as
\begin{align}
\ln P(\Theta | D) = &-0.5 * \Bigg(~\chi^2_{LC} + \chi^2_{RV1} + \chi^2_{RV2} + \Bigg(\frac{T_1-7150~\mathrm{K}}{250~\mathrm{K}} \Bigg)^2\nonumber\\
&+ \Bigg(\frac{T_2-7640~\mathrm{K}}{240~\mathrm{K}} \Bigg)^2 + \Bigg(\frac{e-0.449}{0.005} \Bigg)^2 + \Bigg(\frac{\omega-344.7^\circ}{0.7^\circ} \Bigg)^2\nonumber\\ 
&+ \Bigg(\frac{q-0.96}{0.01} \Bigg)^2 + \Bigg(\frac{\gamma+22.8~\mathrm{km\,s}^{-1}}{0.2~\mathrm{km\,s}^{-1}} \Bigg)^2~\Bigg),
\label{eq:log-likelihood}
\end{align}
where $\chi^2_{LC}$, $\chi^2_{RV1}$, and $\chi^2_{RV2}$ are the respective goodness-of-fit measurements of the light curve, and the primary and secondary RVs, while $T_1$, $T_2$, $e$, $\omega$, $q$, and $\gamma$ are the currently inspected values of the model's primary and secondary effective temperature, eccentricity, longitude of periastron, mass ratio, and systemic velocity, respectively.

At each step the algorithm draws the chains' positions in the parameter space from a uniform, non-informative prior distribution, where we chose 3-$\sigma$ ranges as limits for the parameters that can be constrained from spectroscopy. The limits for the other parameters were chosen by optimising the fit by eye before starting the MCMC computations. Furthermore, we had to introduce a phase shift $\Delta \phi$, to define the zero point in time ($t_0$) as time of periastron passage, since \textsc{phoebe} was designed to model eclipsing binary stars. In a first step, we assumed synchronous rotation for both components and set the synchronicity parameter $F=\omega_\mathrm{rot}/\omega_\mathrm{orb}=1$. Subsequently, we introduced a 2:1 resonance with the orbital period for the primary's rotation period ($F_1=2$), based on our asteroseismic analysis (Sect.~\ref{sec:freqanalysis}), while $F_2$ was allowed to vary. All adjusted parameters and their limits are listed in Table~\ref{tab:mcmc_priors}. Before calculating the $\chi^2$ values, the \textit{Kepler} passband luminosity levels and the bolometric limb-darkening coefficients were computed for the current positions of the chains. The passband limb-darkening coefficients were fixed to values taken from \citet{Claret2011} for models close to our spectroscopic solution.

\begin{table}
\caption{Adjusted parameters and non-informative priors for the MCMC simulation.}
\label{tab:mcmc_priors}
\centering
\begin{tabular}{l c c}
\hline\hline
 & \multicolumn{2}{c}{Orbital parameters} \\
 \hline
 $i_\mathrm{orb}$ (deg) & \multicolumn{2}{c}{60.0 -- 82.5} \\
 $\Delta \phi$ & \multicolumn{2}{c}{0.29252 -- 0.29285} \\
 $e$ & \multicolumn{2}{c}{0.434 -- 0.464} \\
 $\omega$ (deg) & \multicolumn{2}{c}{342.6 -- 346.8} \\
 $a$ ($R_\sun$) & \multicolumn{2}{c}{37.2 -- 44.7} \\
  $q$ & \multicolumn{2}{c}{0.93 -- 0.99} \\
 $\gamma$ (km\,s$^{-1}$) & \multicolumn{2}{c}{$-23.4$ -- $-22.2$} \\
 \hline
 & Primary & Secondary \\
 \hline
 $T_\mathrm{eff}$ (K) & 6400 -- 7900 & 6920 -- 8360 \\
 $\Omega$ & 13.9 -- 19.5 & 15 -- 27 \\
 $F$ & 2 (fixed) & 1.77 -- 1.95 \\
 $\alpha$ & \multicolumn{2}{c}{0.4 -- 1.0} \\
 $\beta$ & \multicolumn{2}{c}{0.0 -- 1.0} \\
 \hline
 \end{tabular}
 \end{table}

One drawback of current binary modelling techniques is that they are computationally intensive and are thus unable to deal with datasets as large as produced by the nominal \textit{Kepler} mission. Therefore, we phase-binned the light curve into 658 bins, reducing the number of data points by a factor of 100. This means fixing the orbital period and time of periastron passage.

\begin{figure*}
\centering
\includegraphics[width=17cm]{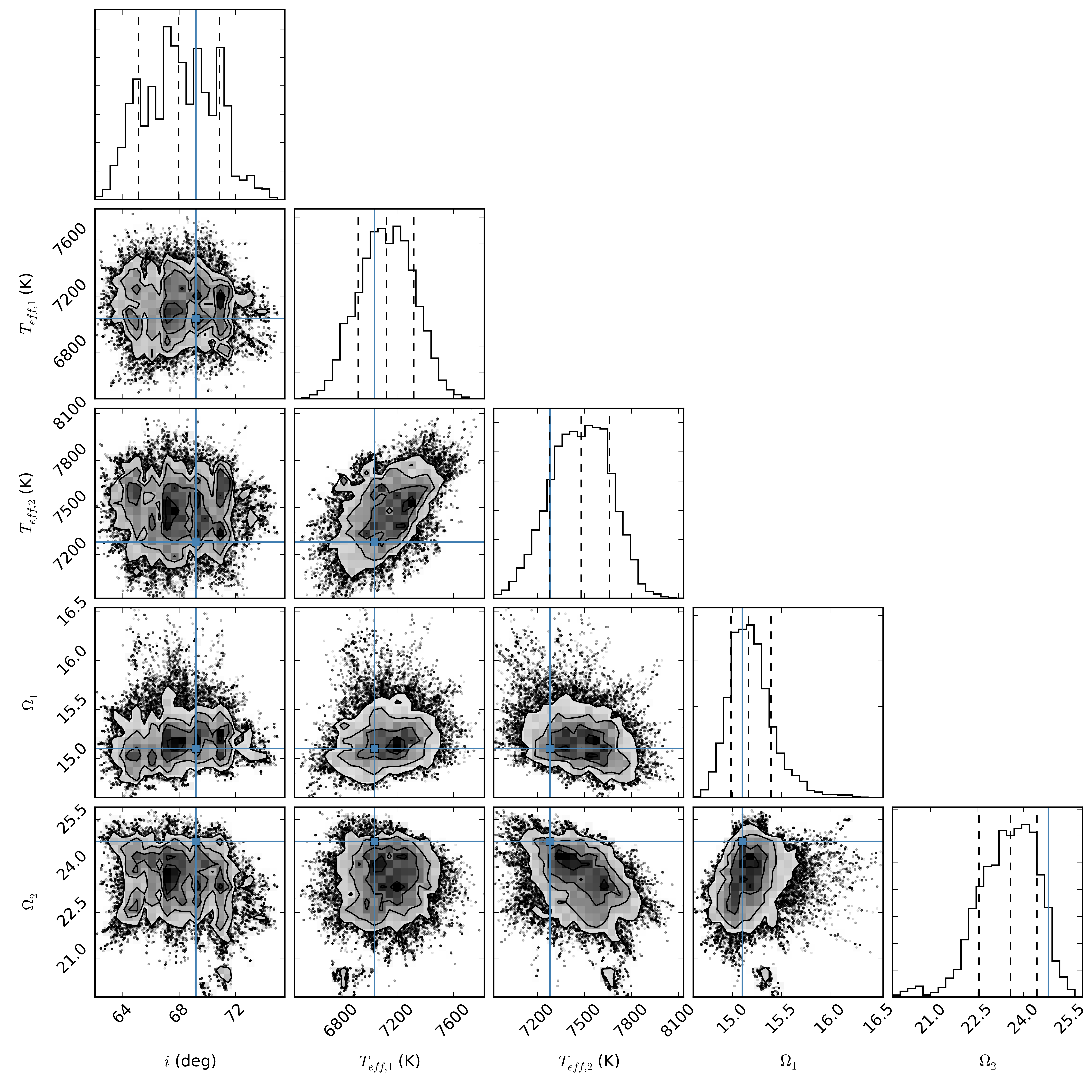}
\caption{Results of fitting the binary light curve, showing the one- and two-dimensional projections of the posterior probability functions of the parameters inclination $i$, primary and secondary effective temperature $T_\mathrm{eff,1,2}$, and primary and secondary surface potential $\Omega_\mathrm{1,2}$. The blue squares are the values of the model with the lowest $\chi^2$. In the 1-d distributions, the 16th, 50th, and 84th percentile are marked by the vertical dashed lines.} 
\label{fig:triangle_teff-pot}
\end{figure*}

The algorithm converged after 2000 iterations of 64 chains and was then reiterated 128\,000 times to explore the parameter space and determine the uncertainties of each parameter. Figure~\ref{fig:triangle_teff-pot} shows the posterior distributions of the following parameters: inclination $i_\mathrm{orb}$, effective temperature $T_\mathrm{eff,1,2}$, and surface potential $\Omega_\mathrm{1,2}$ of the primary and secondary component\footnote{This figure was created using \texttt{triangle.py v0.1.1} \citep[http://dx.doi.org/10.5281/zenodo.11020]{Foreman-Mackey2014}.}. It can be seen that the orbital inclination is not distributed normally, hence the reported uncertainties should be interpreted as a range of possible values. Furthermore, we find no obvious correlations with other parameters. All orbital parameters defined by the RVs are well within the error bars determined from spectral disentangling, while $F_2$ could not be constrained. The result is summarised in Table~\ref{tab:binary} and the best model fit to the binary light curve is shown in Fig.~\ref{fig:lc_fit}, and to the radial velocity curves in Fig.~\ref{fig:RV_fit}.

\subsection{Bolometric albedos and gravity darkening}
\label{sec:proximity}

Other parameters that influence the shape of the light curve are the bolometric albedo $\alpha$ and the gravity darkening exponent $\beta$. The former is linked to reflection and determines the flux fraction that is used to heat the irradiated star, while the latter determines the difference in flux that results from varying surface gravity across the distorted stellar surfaces. The values of these parameters depend on the main energy transport mechanism in the stellar envelope. For predominantly convective envelopes, the theoretically predicted values are $\alpha=0.5$ \citep{Rucinski1989} and $\beta=0.32$ \citep{Lucy1967,Claret2003}, while for radiative envelopes, the commonly used values are $\alpha=\beta=1$ \citep{Rucinski1989,Claret2003}. Both components of KIC\,10080943 lie within in a range of mass and temperature for which convective envelopes are expected, but these are thin and the radiative zone extends almost to the photosphere. We therefore decided not to fix $\alpha_{1,2}$ and $\beta_{1,2}$ for the modelling (Table~\ref{tab:mcmc_priors}).

\begin{figure}
\centering
\includegraphics{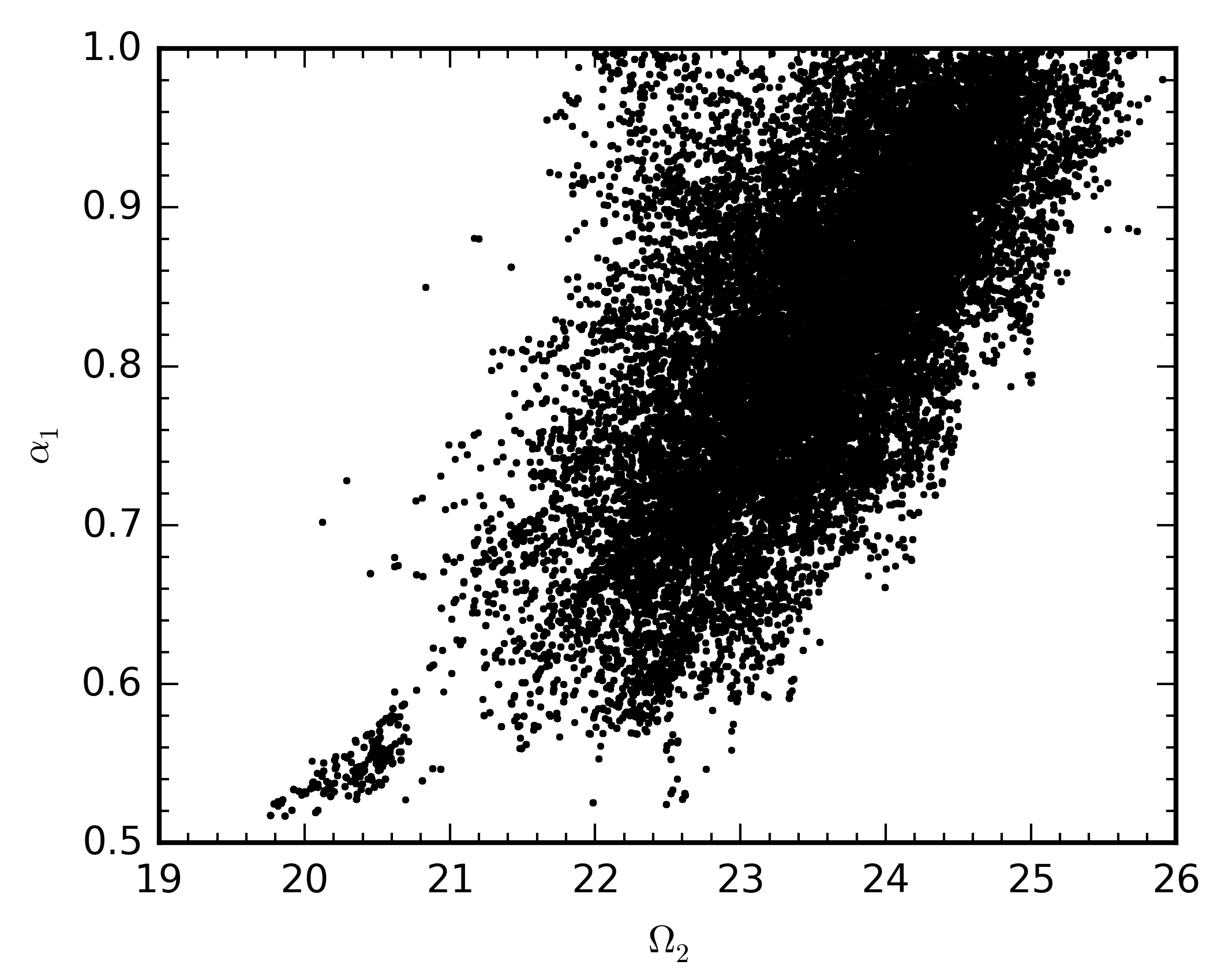}
\caption{Correlation between the secondary surface potential $\Omega_2$ and the primary bolometric albedo $\alpha_1$.}
\label{fig:corr_pot2-alb1}
\end{figure}

We found a weak positive correlation for $\alpha$ and $T_\mathrm{eff}$ for the primary and the secondary and $\alpha_{1}=0.8\pm0.1$, $\alpha_2=0.95\pm0.05$. Between $\Omega_2$ and $\alpha_{2}$, as well as $\alpha_{1}$, there is a stronger correlation (Fig.~\ref{fig:corr_pot2-alb1}), which means that for a smaller secondary radius, a higher albedo is required to fit the observed height of the peak associated with reflection.

\begin{figure}
\centering
\includegraphics{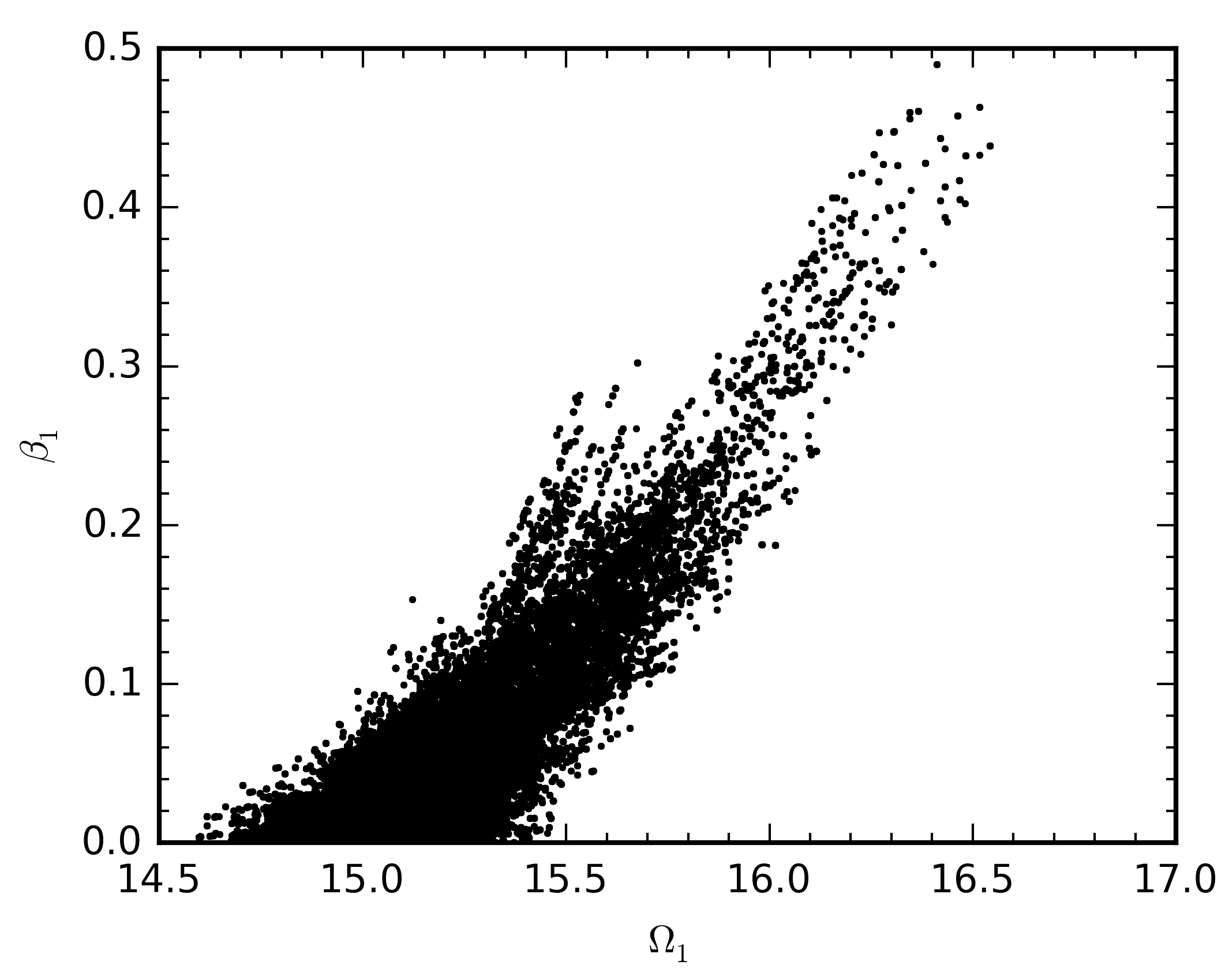}
\caption{Correlation between surface potential $\Omega_1$ and gravity darkening exponent $\beta_1$ of the primary.}
\label{fig:corr_pot1-beta1}
\end{figure}

The gravity-darkening exponents, on the other hand, tend to lower values. In particular for the primary $\beta_1=0.035_{-0.026}^{+0.06}$, while showing a strong correlation with $\Omega_1$ (Fig.~\ref{fig:corr_pot1-beta1}). A similar degeneracy between surface potential and gravity darkening has been reported before by \citet{Beck2014}.

From these degeneracies, we derive the expected surface potentials for the theoretical albedos and gravity-darkening exponents for convective envelopes. Since, any kind of degeneracies represent an uncertainty, we report the surface potentials and radii in Table~\ref{tab:binary} for the full range of possible values, including the MCMC solution. The corresponding ratio $R_1/R_2$ is then in excellent agreement with the spectroscopic value with fixed micro-turbulence.

\section{Pulsation frequency analysis}
\label{sec:freqanalysis}

\begin{figure*}
\centering
\includegraphics{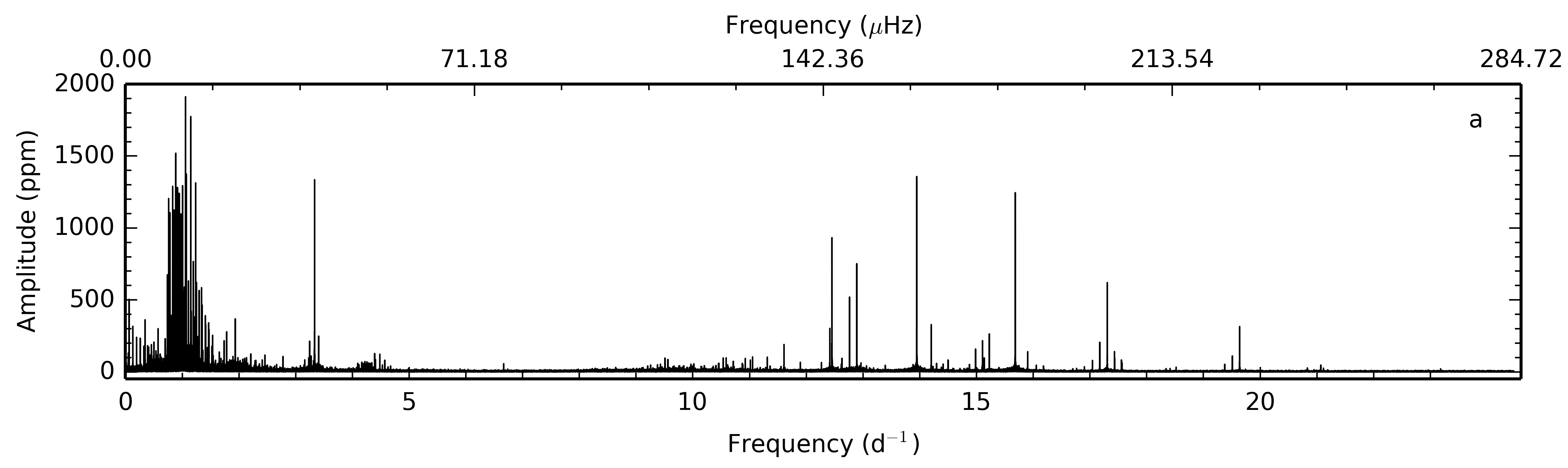}
\includegraphics{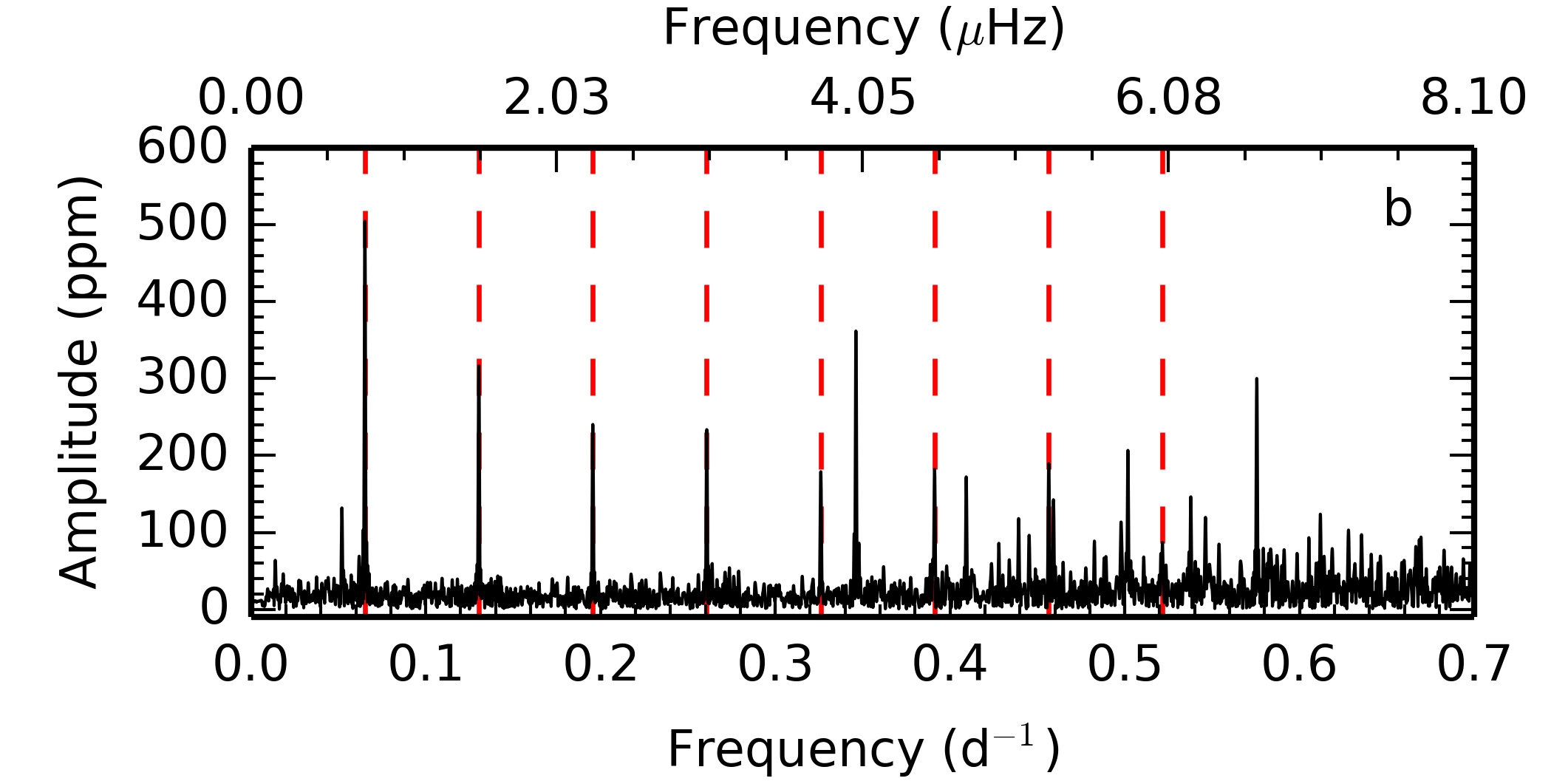}
\includegraphics{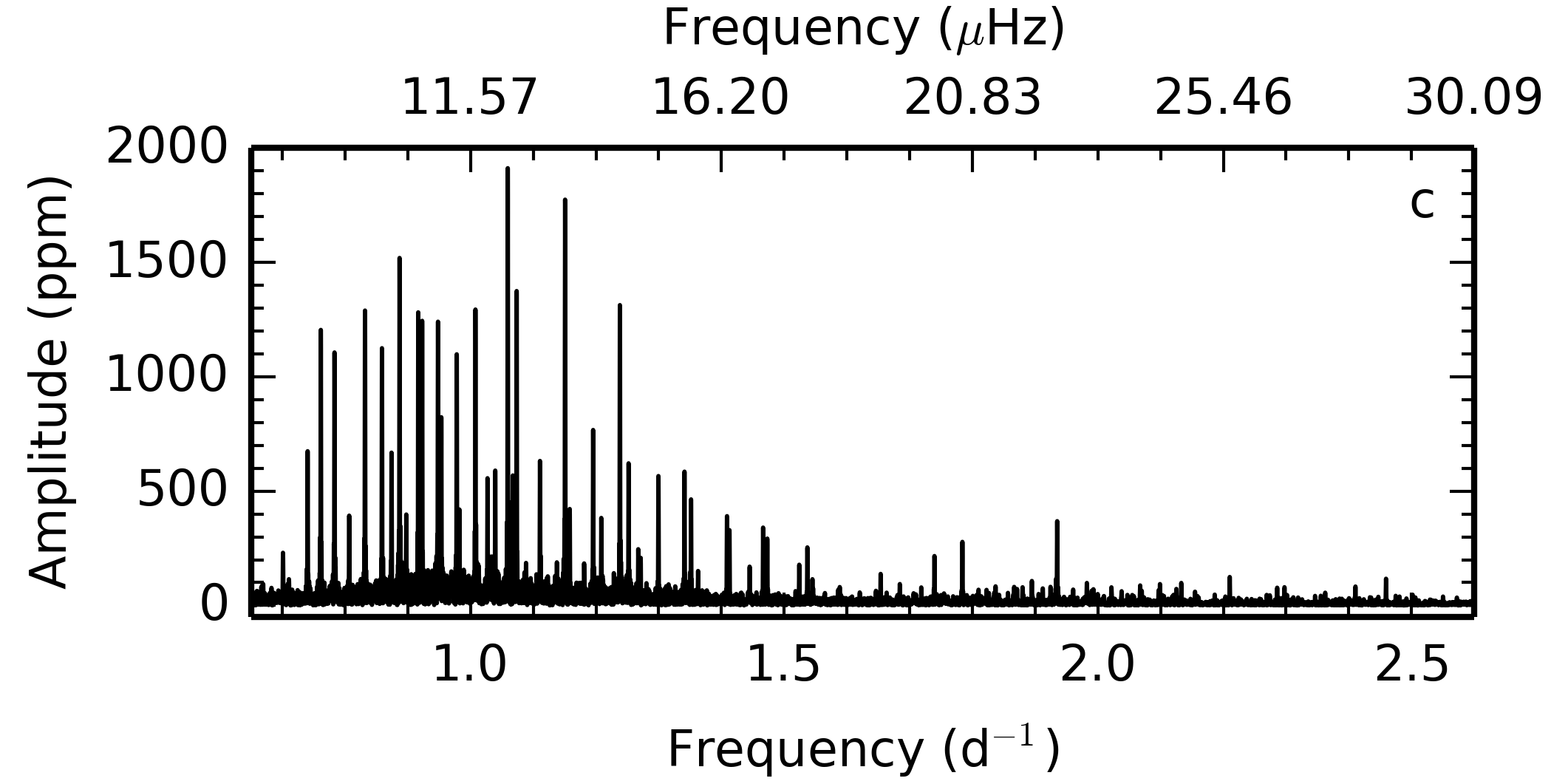}
\includegraphics{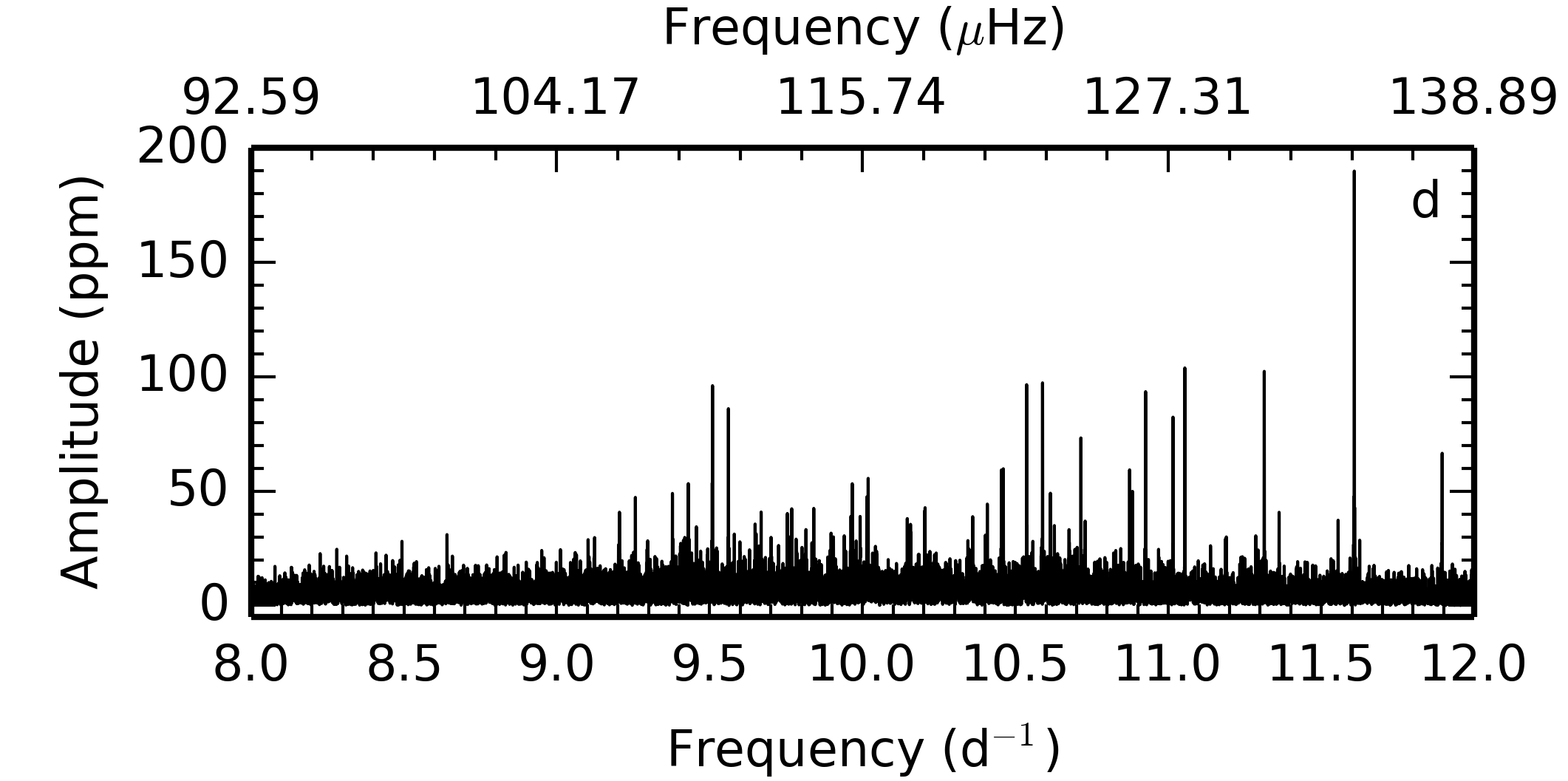}
\includegraphics{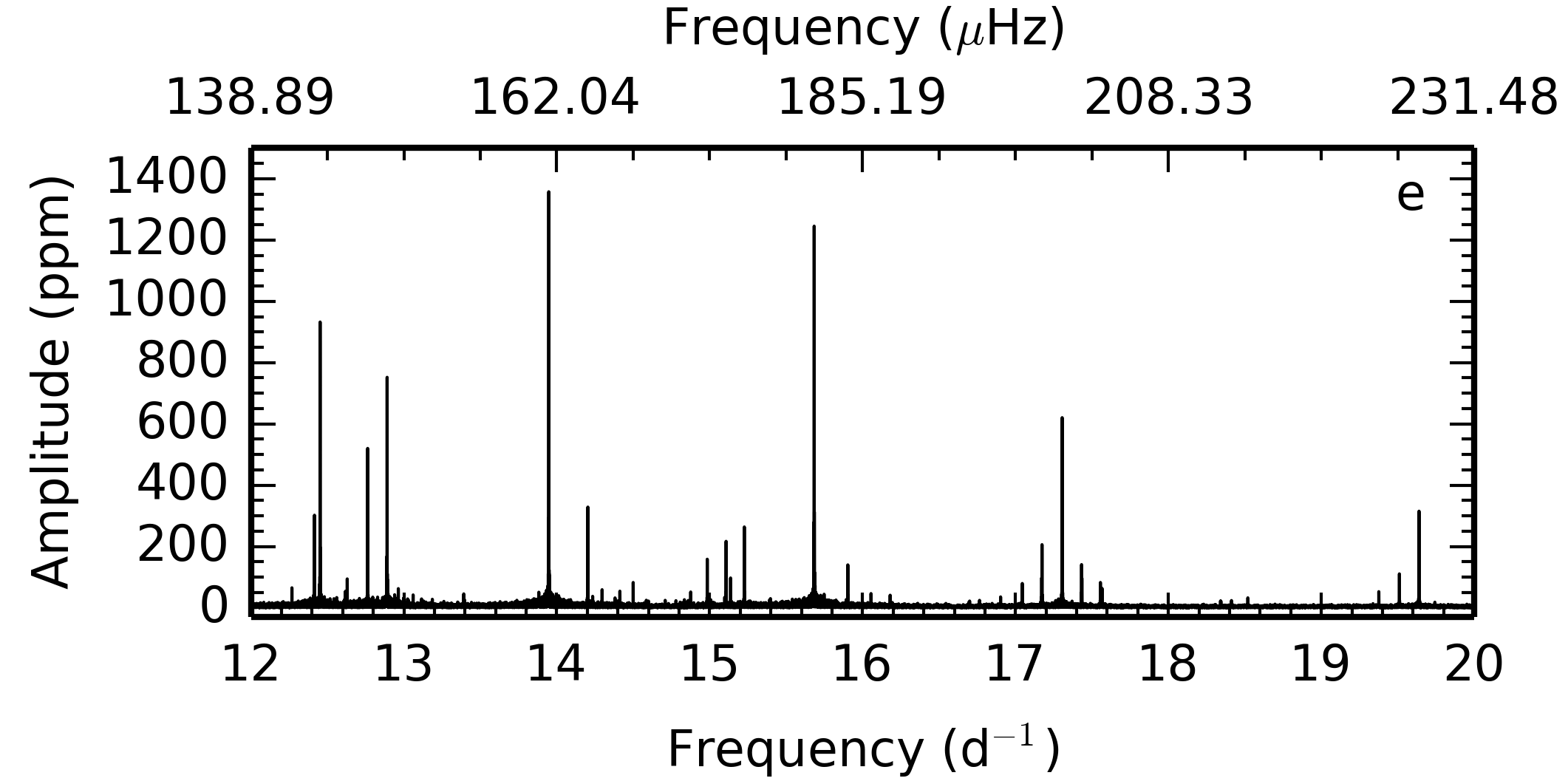}
\caption{\textit{Panel a:} The periodogram of the original light curve of four years of \textit{Kepler} data. \textit{Panel b:} Zoom into the long period range, where the orbital harmonics are indicated by the vertical dashed lines. \textit{Panel c:} Zoom into the g-mode pulsations. \textit{Panel d:} Zoom into the low-amplitude p~modes and into the high-amplitude p~modes (\textit{panel e}).}
\label{fig:pergram}
\end{figure*}

In the \textit{Kepler} light curve, the periodic flux variations caused by stellar pulsations are far more prominent than the binary signal. We extracted this oscillatory behaviour with a Fourier transform \citep[for a detailed description of the method, we refer to][]{Degroote2009}. There are two variability regions clearly distinguishable in the periodogram, displayed in Fig.~\ref{fig:pergram}. Below a frequency of 5~d$^{-1}$, we find pulsation modes of $\gamma$ Dor type, while in the high frequency range, above 8~d$^{-1}$, we find typical $\delta$ Sct pulsation modes. Both regions are highly structured.

As a stop criterion for the iterative prewhitening, we chose a \textit{p}~value below 0.001, meaning there is a 0.1~\% chance that a frequency peak is pure noise. Of all detected peaks, we only considered those with $S/N\geq4$ \citep{Breger1993}, where the noise level was calculated as the mean amplitude of the residual periodogram in a 1~d$^{-1}$ frequency range around each peak. Since the noise level is not uniform across the frequency spectrum, and is higher in the low-frequency regime and in regions where many peaks cluster, an ordinary prewhitening does not reach the high-frequency peaks with low amplitudes ($A<20~$ppm). Hence we applied a Gaussian filter with $\sigma=0.02$~d to the light curve and subtracted the result from the original light curve to treat the high-frequency regime. This method lowers the signal below 9~d$^{-1}$ considerably; the highest peak in this region has $A<110~$ppm. When merging the two sets of the two frequency searches, we therefore selected all peaks $\leq9~\mathrm{d}^{-1}$ from the original light curve and all peaks $>9~\mathrm{d}^{-1}$ from the filtered light curve. We then fitted a harmonic model with these frequencies to the original light curve.

\begin{figure}
\centering
\includegraphics{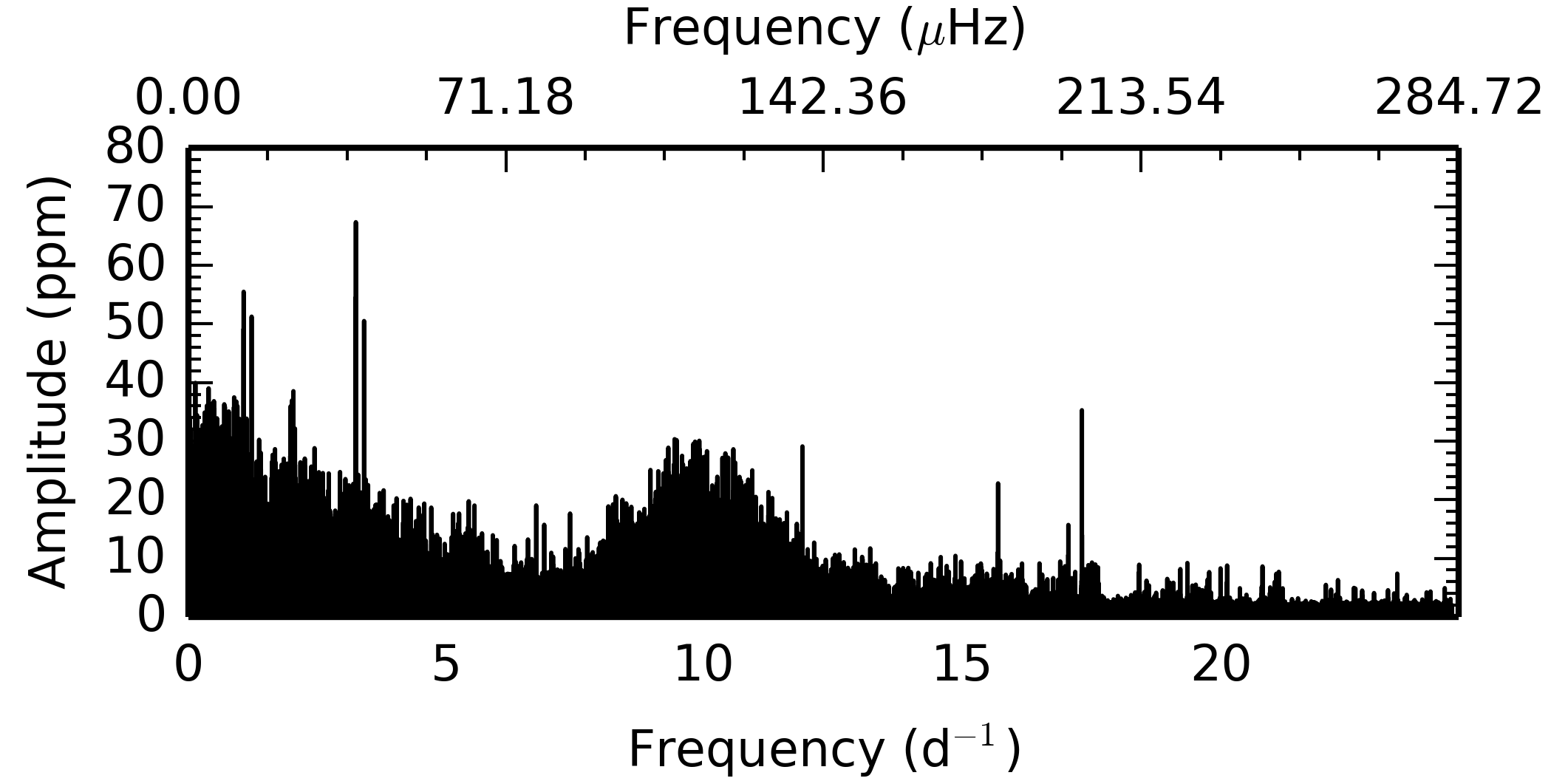}
\caption{Periodogram of the residual light curve after subtracting the binary model and 522 significant frequencies (see text).}
\label{fig:residuals}
\end{figure}

The width of one frequency peak is connected with the \textit{Rayleigh limit} $1/T=0.00068~\mathrm{d}^{-1}$, with $T=1470.4624$~d being the length of the data set. \citet{Loumos1978} found that two peaks with a separation smaller than $2.5/T$ influence each other during the prewhitening process. There are several low-amplitude, yet significant frequencies that are found in the vicinity of high-amplitude peaks with a separation smaller than the Rayleigh limit, which we excluded from the final frequency list. Figure~\ref{fig:residuals} shows the periodogram of the residual light curve, after subtracting the binary model calculated in Sect.~\ref{sec:binary} and 522 significant frequencies.

We calculated the errors on the frequency, amplitude, and phase, following the description of \citet{Montgomery1999}. These errors rely on the assumption of white noise in the data and therefore underestimate the real error, in this current study by a factor $\sim2$ owing to correlation effects \citep{Schwarzenberg-Czerny2003}. We applied this correction factor to the errors. All parameters, their correlation-corrected uncertainties, and $S/N$ values are listed in Table~\ref{tab:fourier_results}.

As already mentioned in Sect.~\ref{sec:binary}, we also detect the binary ellipsoidal variation and reflection signal as orbital harmonics in the periodogram (see Fig.~\ref{fig:pergram}, panel b). To disentangle both signals, we have to perform an iterative analysis. After subtracting the binary model, we only found minimal changes in the determined pulsation frequencies, in comparison with the solution derived from the original light curve with the orbital harmonics prewhitened in Fourier space rather than the binary light curve in the time domain. The frequency $2.2940~\mathrm{d}^{-1}$ with $S/N=4.01$ is not detected anymore when working with the residual light curve after orbital subtraction. Six additional frequencies were detected: $f_{255}$, $f_{256}$, $f_{265}$, $f_{268}$, $f_{276}$, and $f_{512}$.

\subsection{Tidally influenced pulsations}
\label{sec:tidalpuls}

Another source for frequencies at multiples of the binary orbit are tidally induced or influenced pulsations. Examples of pulsating stars in eccentric binaries exhibiting tidally excited modes are given by \citet{Welsh2011} and \citet{Hambleton2013}, where almost all g~modes could be attributed to this driving mechanism. In the case of KIC\,10080943, we only see weak tidal influence on the pulsations. Altogether we found four frequencies that are in the vicinity of integer multiples of the orbital frequency, listed in Table~\ref{tab:tidalpuls}, where the relative frequency $f_r = f_{obs}-nf_{orb}>1/T$.

The frequency $f_{128}=0.45663\approx7f_{orb}$ coincides with the orbital harmonics caused by the non-sinusoidal binary signal. Consequently, it was prewhitened together with the orbital harmonics to study the pulsations alone in the first iteration of the binary-pulsation analysis. By cutting the brightening from the pulsation-residual light curve and fitting a polynomial to the phases outside the brightening, we manually separated the two signals. All non-sinusoidal signals at the orbital period are removed, while oscillations at orbital harmonics remain, creating a peak in the frequency spectrum. When a binary model was subtracted in subsequent iterations, as opposed to orbital harmonics, this confusion no longer arose.

\begin{table}
\caption{Frequencies and amplitudes close to multiples of the orbital frequency are given, as well as the frequency ratio $f/f_\mathrm{orb}$.}
\label{tab:tidalpuls}
\centering
\begin{tabular}{l l r r}
\hline\hline
 & \multicolumn{1}{c}{$f$} & \multicolumn{1}{c}{$A$} & \multicolumn{1}{c}{$f/f_\mathrm{orb}$} \\
 & \multicolumn{1}{c}{d$^{-1}$} & \multicolumn{1}{c}{ppm} & \\
\hline
$f_{16}$ & 0.978056 & 1131.2 & 15.00084 \\
$f_{128}$ & 0.45663 & 75.2 & 7.00353 \\
$f_{147}$ & 1.56487 & 71.9 & 24.00107 \\
$f_{388}$ & 2.6734 & 28.9 & 41.00307 \\
\hline
\end{tabular}
\end{table}

\subsection{The g modes}
\label{sec:periodspacing}

The g~modes (Fig.~\ref{fig:pergram}, panel c) in KIC\,10080943 are discussed by \citet{Keen2015}, who performed an independent analysis. They found six series of modes equally spaced in period. Five of these consist of $\ell=1$ dipole modes, which form rotationally split mutliplets, a series of doublets, and a series of triplets. This indicates that both stars pulsate in g~modes. Since, for the doublets the central $m=0$ modes cannot be detected, the frequency splitting $\Delta f_\mathrm{g1}\approx0.091~\mathrm{d}^{-1}$ of the outer components is twice the actual rotational splitting. Approximating the Ledoux constant for dipole g~modes by $C_{n,\ell}\approx0.5$ then leads to a rotational period $P_\mathrm{rot,g1}\approx11$~d. For the triplets, on the other hand, all components are detected and they are split by $\Delta f_\mathrm{g2}\approx0.07~\mathrm{d}^{-1}$, yielding $P_\mathrm{rot,g2}\approx7$~d. Additionally, they found evidence for modes of $\ell=2$. For figures and tables of the period spacing and rotational splitting in the g~modes, we refer to \citet{Keen2015}. In the analysis performed for this paper, we confirm their results.

\subsection{The p modes}
\label{sec:frequencysplitting}

\begin{figure*}
\centering
\includegraphics{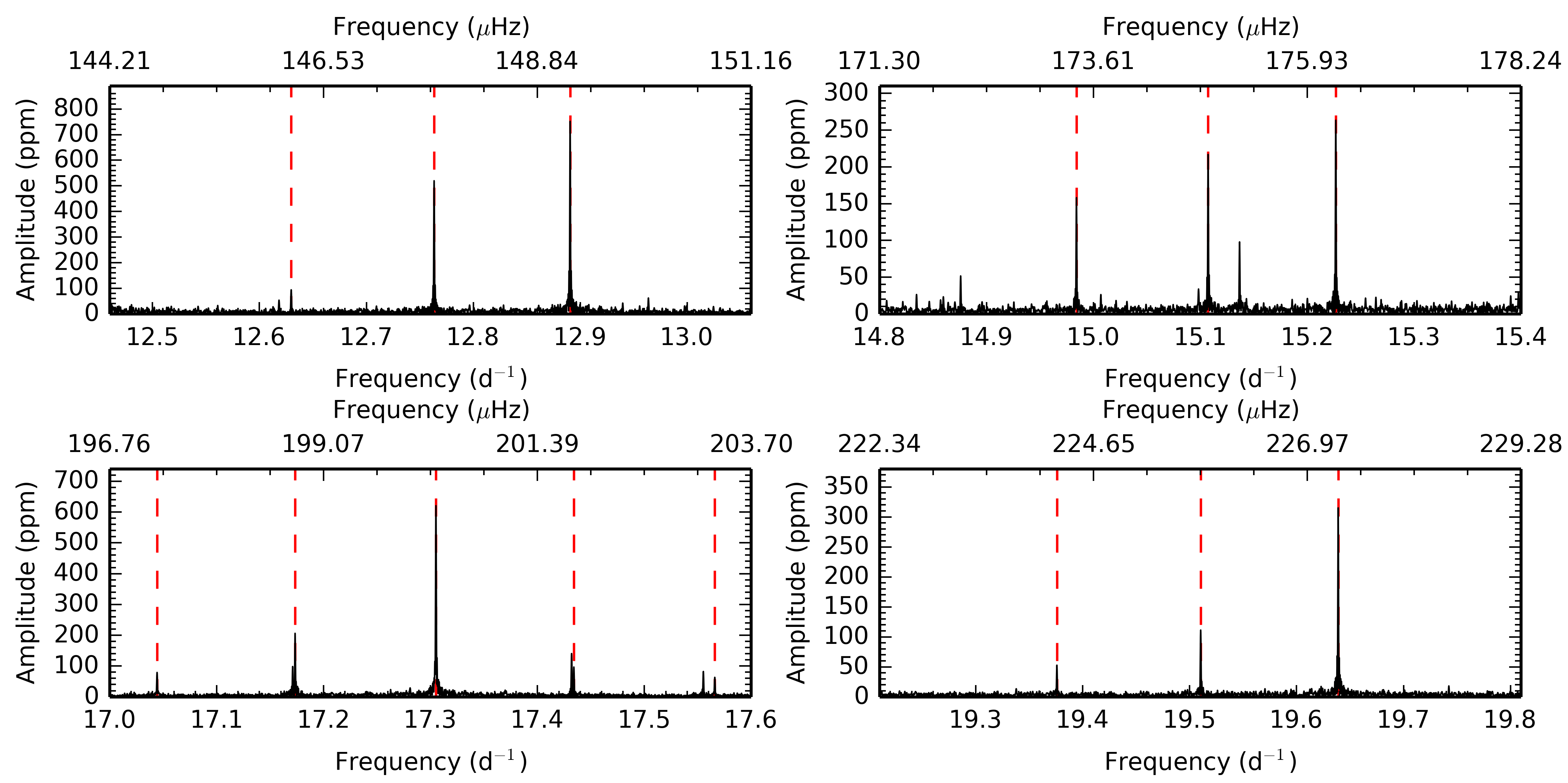}
\caption{Frequency splittings in the p-mode regime of the periodogram. The vertical dashed lines indicate the modes, composing the multiplets. The frequencies, amplitudes, and frequency differences are listed in Table~\ref{tab:frequencysplitting}. \textit{Upper left:} Triplet with mean separation $\Delta f_1=0.1304\pm0.0032~\mathrm{d}^{-1}$. \textit{Upper right:} Triplet with a mean separation $\Delta f_2=0.1213\pm0.0019~\mathrm{d}^{-1}$. \textit{Lower left:} Quintuplet with mean separation $\Delta f_1=0.1304\pm0.0013~\mathrm{d}^{-1}$. \textit{Lower right:} Triplet with mean separation $\Delta f=0.1315\pm0.0029~\mathrm{d}^{-1}$.}
\label{fig:frequencysplitting}
\end{figure*}

In the highly structured p-mode regime ($f>8~\mathrm{d}^{-1}$; Fig.~\ref{fig:pergram}, panels d and e) we detected four multiplets caused by rotational splitting, which are displayed in Fig.~\ref{fig:frequencysplitting} and listed in Table~\ref{tab:frequencysplitting}. The most prominent feature is the quintuplet around 17.3~d$^{-1}$, which has a mean frequency splitting of $\Delta f_\mathrm{p1}=0.1304\pm0.0013~\mathrm{d}^{-1}=2 f_\mathrm{orb}$. In the triplet around 15~d$^{-1}$, we found a smaller splitting value of $\Delta f_\mathrm{p2} = 0.1213\pm0.0019~\mathrm{d}^{-1}$. Furthermore, none of the multiplets are symmetric about their centre frequency, which could be indicative of second-order rotational effects \citep[e.g.][]{Saio1981} or magnetic splitting, in addition to rotational splitting \citep[e.g.][]{Goode1992}. This is most noticeable for the triplet at 19.5~d$^{-1}$\,, where the mean splitting value ($\Delta f = 0.1315\pm0.0029~\mathrm{d}^{-1}$) is also larger, although within the error bars of $\Delta f_\mathrm{p1}$. We note that the rotational splitting values that occur in the p~modes are different from the ones found in the g~modes, which will be further discussed in Sect.~\ref{sec:discussion}.

Furthermore, we find that the two highest-amplitude peaks $f_4=13.94759~\mathrm{d}^{-1}$ and $f_{13}=15.683330~\mathrm{d}^{-1}$ are singlets, and hence likely to be radial modes, although their degree as well as radial order remain unidentified at this stage. In Sect.~\ref{sec:PM} we present evidence that these two frequencies originate in the two different stars.

Another spacing, which occurs several times among frequency pairs, is $\Delta f=0.052~\mathrm{d}^{-1}$. It is most striking in the frequency range $8~\mathrm{d}^{-1}< f < 12~\mathrm{d}^{-1}$ (Fig.~\ref{fig:pergram}, panel d) and is connected to the difference of two high-amplitude g modes $f_1-f_{12}=0.05152\pm0.000013$. We interpret these p-mode frequency pairs as due to non-linear resonant mode coupling. The spacing also appears in the periodogram as $f_{87}=0.05208~\mathrm{d}^{-1}$. A similar effect has been observed in the slowly pulsating B-type (SPB) binary KIC\,6352430 by \citet{Papics2013}, who give a similar interpretation.

\begin{table}
\caption{Frequency splittings in the p mode regime.}
\label{tab:frequencysplitting}
\centering
\begin{tabular}{l r l}
\hline\hline
\multicolumn{1}{c}{$f$} & \multicolumn{1}{c}{$A$} & \multicolumn{1}{c}{$\Delta f$} \\
\multicolumn{1}{c}{d$^{-1}$} & \multicolumn{1}{c}{ppm} & \multicolumn{1}{c}{d$^{-1}$} \\
\hline
12.62972 & 95.1 & \\
12.763339 & 533.5 & $0.133619\pm0.000020$ \\
12.890541 & 758.4 & $0.127202\pm0.000005$ \\
& & \\
14.98412 & 160.9 & \\
15.10724 & 220.1 & $0.12312\pm0.000014$ \\
15.226646 & 257.2 & $0.119406\pm0.000013$ \\
& & \\
17.04422 & 83.0 & \\
17.17332 & 201.9 & $0.12910\pm0.000032$ \\
17.305043 & 616.1 & $0.131723\pm0.000011$ \\
17.43413 & 108.4 & $0.129087\pm0.000020$ \\
17.56586 & 64.4 & $0.13173\pm0.000045$ \\
& & \\
19.37580 & 57.3 & \\
19.51028 & 115.4 & $0.13448\pm0.000045$ \\
19.638879 & 315.5 & $0.1285994\pm0.000021$ \\
\hline
\end{tabular}
\end{table}

\subsection{Combination frequencies}

Non-linear pulsation effects can manifest themselves in the Fourier spectrum as combination frequencies and harmonics of a few parent frequencies. \citet{Papics2015} and \citet{Kurtz2015} have recently used them to explain the skewed, non-sinusoidal light curves of many $\gamma$\,Dor and SPB stars. We also identify non-linearity in the light curve of KIC\,10080943 from the detection of several low-order combination frequencies of the form $nf_i\pm mf_j = f_k$, where $n$ and $m$ are small integers, $f_i$ and $f_j$ are called the parent frequencies, and $f_k$ the combination frequency. To keep the number of combinations manageable, we limit the search for combinations to $(n+m)\leq2$ and to parent frequencies among the 40 highest amplitude modes. We note that from those 40 modes, only five are not identified to be part of a period spacing series or rotational multiplet. Owing to the dense frequency spectrum, there are several combinations that can be made by pure chance \citep{Papics2012a}, and several different parents can combine to the same combination frequency. Nevertheless, we found clear and obvious patterns.

Frequencies $f_4=13.94759~\mathrm{d}^{-1}$ and $f_8=3.33350~\mathrm{d}^{-1}$ create combinations (sums and differences) with almost all mode frequencies of the rotationally split doublets in the g modes. The frequency group between 4~d$^{-1}$ and 4.7~d$^{-1}$ is fully explained through resonances with $f_8$, as discussed by \citet{Keen2015}. Moreover, the single peak between the g- and p-mode regions can be related to $2 f_8=6.66697~\mathrm{d}^{-1}$ (see Fig.~\ref{fig:pergram}, panel a).

Nonlinear mode coupling also happens among the high-amplitude g~modes, which (almost exclusively) explain the frequencies between 1.6~d$^{-1}$ and 2.6~d$^{-1}$. Although some combinations are created by modes of the frequency triplets, by far the majority of combinations are caused by the frequency doublets, due to their higher amplitudes.

\subsection{Phase modulation}
\label{sec:PM}

The motion of the stars in the binary orbit causes a periodic variation in the path length travelled by the light from the stars to the Earth, which results in a phase change of the observed pulsation over the orbital period \citep{Murphy2014}. Since these changes should be in anti-phase for the two components, we used this method to determine which pulsation frequencies originate in which star.

The entire \textit{Kepler} light curve was divided into 5-day segments to retrieve the phase variations, and the phases of the pulsations were determined with the frequencies fixed to the values in Table~\ref{tab:fourier_results}. These phase variations were then converted into light arrival time delays. A Fourier transform of these time delays gives the orbital period, the projected light travel time across the orbit, and the eccentricity \citep[a full description is provided by][]{Murphy2014,Murphy2015}. For a period as short as 15.3364 d, the segment size is a trade off between the frequency resolution for each segment and the sampling rate of the orbit. This is easier to achieve for the highest-amplitude p~modes. Despite the obvious limitation of this method for short-period binaries, we were able to determine the phases of the orbital variations of seven modes (excluding $f_{18}$, which has strong beating with a close frequency), given in Table~\ref{tab:PM}. For $f_4$ and $f_{13}$ the phases are $2.06\pm0.08$ and $5.31\pm0.17$, respectively, which are $\pi$ rad out of phase within the errors. They have time delays of $44.0\pm3.5$ s and $36.2\pm4.8$ s, respectively, where the smallest time delay belongs to the star with the smaller $a\sin i_\mathrm{orb}$ and larger mass. Consequently, the peaks whose orbital variation has a phase near 5.0 belong to the primary, while those with phases near 2.0 belong to the secondary.

This method can also be used to derive orbital parameters, without the use of ground-based spectroscopy. For KIC\,10080943 they are in good agreement with our results derived from RVs and spectral disentangling, but the results from phase modulation (PM) have much larger error bars, due to the short orbital period and a segment size of only 5 days.

Another effect caused by the light travel time delay is modulation of frequencies, which creates side lobes at the orbital frequency in the periodogram. We detected these side lobes for the high-amplitude p~modes. These can also be used to study an orbit photometrically \citep[e.g.][]{Shibahashi2012,Telting2014,Shibahashi2015}, but once again, our spectroscopy is far superior for deducing orbital properties in the case of KIC\,10080943.

\begin{table}
\centering
\caption{The frequencies and orbital variation phases for 7 of the 8 highest amplitude p~modes.}
\label{tab:PM}
\begin{tabular}{c c c c}
\hline
\hline
 \multicolumn{2}{c}{Frequency} & Orbital phase & Star\\
& d$^{-1}$ & rad &\\
\hline
$f_4 $&$ 13.947586 $&$ 2.06\pm0.08$ & 2\\
$f_{13} $&$ 15.683330 $&$ 5.31\pm0.17$ & 1\\
$f_{21} $&$ 12.890541 $&$ 5.75\pm0.19$ & 1\\
$f_{27} $&$ 17.305043 $&$ 3.64\pm0.26$ & ?\\
$f_{32} $&$ 12.763339 $&$ 0.87\pm0.24$ & ?\\
$f_{45} $&$ 14.203334 $&$ 2.03\pm0.34$ & 2\\
$f_{46} $&$ 19.638879 $&$ 4.75\pm0.28$ & 1\\
\hline
\end{tabular}
\end{table}

\section{Discussion}
\label{sec:discussion}

\subsection{Core-to-surface rotation}
\citet{Keen2015} detected two different frequency-splitting values in the g~modes of KIC\,10080943 from analysis of the \textit{Kepler} light curve without orbital prewhitening. In this paper we confirm this result after subtraction of the binary signal from the light curve followed by pulsation frequency analysis. Moreover, we find two different splitting values in the p~modes as well. This is strong evidence that both stars pulsate as p- and g-mode hybrids. Our binary analysis yields similar masses for both stars, which supports this hypothesis. By detecting rotational splitting in the g~modes, which are sensitive to the region close to the stellar core, as well as in the p~modes, which have their highest amplitudes in the stellar envelope, we can constrain the internal rotation profile of a star in a largely model-independent way, as explained in detail in \citet{Kurtz2014} and in \citet{Saio2015} for two previously analysed stars with similar mass. Here, we achieve this via the detected frequency patterns and combination frequencies. The p-mode frequency $f_4$ only creates combinations with the frequencies of the g-mode doublets, while the frequency $f_{13}=15.68333~\mathrm{d}^{-1}$ does not create combination frequencies in this way and also does not originate in the same star as $f_4$ (Table~\ref{tab:PM}). The PM results further show that $f_{21}=12.890541~\mathrm{d}^{-1}$, which is part of a triplet with rotational splitting $\Delta f_\mathrm{p1}=0.1304~\mathrm{d}^{-1}$, does stem from the same star as $f_4$. We therefore conclude that the p-mode splitting that corresponds to the rotational period $P_\mathrm{rot,g1} \approx 11~\mathrm{d}$ of the g-mode doublet has to be $\Delta f_\mathrm{p2} = 0.1213~\mathrm{d}^{-1}$. For pure p~modes, the Ledoux constant $C_{n,\ell}\approx0$ and $\beta_{n,\ell}=1-C_{n,\ell}\approx 1$, resulting in a rotational period $P_\mathrm{rot,p2}\approx8.2$~d. Thus, we conclude a core-to-surface rotation rate for the secondary of approximately $P_\mathrm{core,2}/P_\mathrm{surface,2} \approx1.3$. Similarly for the primary, $P_\mathrm{rot,p1}\approx7.7~\mathrm{d}$ and $P_\mathrm{core,1}/P_\mathrm{surface,1} \approx 0.9$.

The core-to-surface rotation rates derived here are rough first estimates, as the proper computation of the p-mode Ledoux constant requires seismic modelling of the zonal modes, taking into account a possible influence by the g-mode cavity, as well as the observed departure from equidistance in the p-mode splittings (Table~\ref{tab:frequencysplitting}). Also, the g-mode splittings vary with frequency, which \citet{Keen2015} attributed to the dependence of the Ledoux constant on radial order $n$. \citet{Triana2015} deduced the internal rotation profile of the SPB KIC\,10526294 from g-mode frequency inversion, after detailed seismic modelling of the zonal modes by \citet{Moravveji2015}. We plan to follow similar methodology, as in these two papers, for our future detailed modelling of KIC\,10080943, with the aim of determining the interior rotation profiles of its components with better precision than the first rough estimates we present here.

\subsection{Spin-orbit alignment}

It is usually assumed that the rotational axis of binary components are aligned with the axis of the orbital plane. The low amplitudes of the zonal g~modes suggest that the stars are seen under a high inclination angle \citep{Keen2015}, whereas the configuration of the p-mode multiplets seem to indicate a low inclination angle. Since heat-driven pulsations might not be excited to similar intrinsic amplitudes, the splitting geometry only provides weak evidence for the inclination of the stellar pulsation axis. The orbital inclination angle has to be below 82.5$^\circ$, given that no eclipses are observed. We can further test the assumption $i_\mathrm{orb} \simeq i_\mathrm{rot}$ by deriving a radius with the rotational splitting and $v\sin i_\mathrm{rot}$ for $i_\mathrm{orb}=68\pm3^{\degr}$. For the primary ($f_\mathrm{surface,1}=0.1304\pm0.0013~\mathrm{d}^{-1}$, $v_1 \sin i_\mathrm{rot,1}=18.7\pm1.2~\mathrm{km\;s}^{-1}$), we find a radius of $3.06\pm0.22~R_\sun$, and for the secondary ($f_\mathrm{surface,2}=0.1213\pm0.0019~\mathrm{d}^{-1}$, $v_2 \sin i_\mathrm{rot,2}=13.8\pm1.6~\mathrm{km\,s}^{-1}$) $2.43\pm0.29~R_\sun$, giving a ratio $R_1/R_2=1.26\pm0.18$. Using the $v \sin i_\mathrm{rot}$ values from the spectroscopic solution with micro-turbulence as a free parameter makes marginal difference and gives a ratio $R_1/R_2=1.24\pm0.17$. The values derived here are in agreement with the radii derived from binary modelling and spectroscopic analysis, given the 1-$\sigma$ uncertainties in Table~\ref{tab:binary}, as well as the assumed spin-orbit alignment.

\section{Summary}
\label{sec:summary}

From the extraordinarily rich frequency spectrum of the four-year \textit{Kepler} light curve of KIC\,10080943, we have shown that this eccentric binary system contains two $\gamma$\,Dor/$\delta$\,Sct hybrids. The analysis of the clearly separated g-mode range, presented by \citet{Keen2015} shows the discovery of six period spacing series, which form rotationally split triplets of prograde, zonal, and retrograde modes for one star, and doublets of prograde and retrograde modes for the other star. In this paper, we find rotationally split multiplets with two distinct spacings in the p-mode range, where one star seems to have a 2:1 resonance with the orbital period. Using combination frequencies we manage to link period spacing in the g~modes with rotational splitting in the p~modes for both components. We derive that the primary has a core that rotates slightly more rapidly than its surface, while the secondary's core rotates more slowly than its surface. Similar results have been obtained by \citet{Kurtz2014} and \citet{Saio2015} for two single hybrid g- and p-mode pulsators. The phase modulation detected for the high-amplitude p~modes of KIC\,10080943, which is caused by the light travel time effect in the binary orbit, suggests that the secondary is the stronger pulsator.

Buried beneath the high amplitude pulsations, we discover ellipsoidal variation and reflection with an amplitude of only $0.2\%$. The effects are not dominating the light curve, since the 15.3364-d orbit with $e=0.449$ results in a separation of $\sim22.6~R_\sun$ at periastron. We derived the orbital solution from 26 ground-based spectra, showing lines of both components with a mass ratio close to unity. From the disentangled spectra, we confirmed the early F spectral type for both stars, with temperatures $T_\mathrm{eff,1}=7150\pm250$ K and $T_\mathrm{eff,2}=7640\pm240$ K. The surface gravities can only be moderately constrained with spectroscopy. A high-precision estimate was, however, achieved by the binary fit obtained for the light curve. We used \textsc{phoebe} to compute the models and an MCMC simulation to improve the fit, as well as derive the posterior probability distribution for each parameter. The inclination angle is not normally distributed and a range of possible values is $i_\mathrm{orb}=68^{\degr}\pm3^{\degr}$. For this range, we derive absolute masses $M_1=2.0\pm0.1~M_\sun$ and $M_2=1.9\pm0.1~M_\sun$. From the fitted surface potentials, corrected for degeneracies with albedo and gravity darkening, we find the radii $R_1=2.9\pm0.1~R_\sun$ and $R_2=2.1\pm0.2~R_\sun$. From these values we derive surface gravities $\log g_1 = 3.81\pm0.03$ and $\log g_2 = 4.1\pm0.1$. The example presented in this paper illustrates that, for non-eclipsing binaries, it is necessary to combine spectroscopic, photometric, and asteroseismic analyses to obtain a good characterisation of the system. Only double-lined eclipsing binaries allow us to derive high-precision fundamental parameters in a largely model-independent way, and without suffering from degeneracies in the same way as binaries with only a brightening signal.

 As stars evolve along the main sequence, they develop a chemical composition gradient near the core, which creates a periodic signal in the period spacing in the $\Delta p$ vs. $p$ diagram \citep{VanReeth2015b}. We can use the period spacing series detected for KIC\,10080943 to derive the properties of the mixing mechanisms taking place in the deep stellar interior. Further crucial constraints on the stellar structure of both components of KIC\,10080943, keeping in mind its close binary nature and accompanying tidal forces, can be derived from the rotation rates that are detected for the region near the cores and the surfaces. Tight asteroseismic constraints on the interior rotation profile of both stars can only be drawn from the detailed comparison to stellar models, which will be subject of a subsequent paper.

\begin{acknowledgements}
Part of the research included in this manuscript was based on funding from the Research Council of KU\,Leuven, Belgium, under grant GOA/2013/012, from the European Community's Seventh Framework Programme FP7-SPACE-2011-1, project number 312844 (SpaceInn), and from the Fund for Scientific Research of Flanders (FWO), under grant agreement G.0B69.13. This research was supported by the Australian Research Council. Funding for the Stellar Astrophysics Centre is provided by the Danish National Research Foundation (grant agreement no.: DNRF106). VSS, CA, SB, and PIP are grateful to the Kavli Institute of Theoretical Physics at the University of California, Santa Barbara, for the kind hospitality during the research programme, ``Galactic Archaeology and Precision Stellar Astrophysics'', during which part of the present research was conducted. Funding for the \textit{Kepler} mission is provided by the NASA Science Mission Directorate. The authors wish to thank the entire \textit{Kepler} team for all their efforts.
\end{acknowledgements}

\bibliographystyle{aa} 
\bibliography{bibliography}

\begin{thebibliography}{76}
\expandafter\ifx\csname natexlab\endcsname\relax\def\natexlab#1{#1}\fi

\bibitem[{{Aerts}(2015)}]{Aerts2015}
{Aerts}, C. 2015, Astronomische Nachrichten, 336, 477

\bibitem[{{Auvergne} {et~al.}(2009){Auvergne}, {Bodin}, {Boisnard}, {Buey},
  {Chaintreuil}, {Epstein}, {Jouret}, {Lam-Trong}, {Levacher}, {Magnan},
  {Perez}, {Plasson}, {Plesseria}, {Peter}, {Steller}, {Tiph{\`e}ne}, {Baglin},
  {Agogu{\'e}}, {Appourchaux}, {Barbet}, {Beaufort}, {Bellenger}, {Berlin},
  {Bernardi}, {Blouin}, {Boumier}, {Bonneau}, {Briet}, {Butler}, {Cautain},
  {Chiavassa}, {Costes}, {Cuvilho}, {Cunha-Parro}, {de Oliveira Fialho},
  {Decaudin}, {Defise}, {Djalal}, {Docclo}, {Drummond}, {Dupuis}, {Exil},
  {Faur{\'e}}, {Gaboriaud}, {Gamet}, {Gavalda}, {Grolleau}, {Gueguen},
  {Guivarc'h}, {Guterman}, {Hasiba}, {Huntzinger}, {Hustaix}, {Imbert},
  {Jeanville}, {Johlander}, {Jorda}, {Journoud}, {Karioty}, {Kerjean},
  {Lafond}, {Lapeyrere}, {Landiech}, {Larqu{\'e}}, {Laudet}, {Le Merrer},
  {Leporati}, {Leruyet}, {Levieuge}, {Llebaria}, {Martin}, {Mazy}, {Mesnager},
  {Michel}, {Moalic}, {Monjoin}, {Naudet}, {Neukirchner}, {Nguyen-Kim},
  {Ollivier}, {Orcesi}, {Ottacher}, {Oulali}, {Parisot}, {Perruchot},
  {Piacentino}, {Pinheiro da Silva}, {Platzer}, {Pontet}, {Pradines},
  {Quentin}, {Rohbeck}, {Rolland}, {Rollenhagen}, {Romagnan}, {Russ}, {Samadi},
  {Schmidt}, {Schwartz}, {Sebbag}, {Smit}, {Sunter}, {Tello}, {Toulouse},
  {Ulmer}, {Vandermarcq}, {Vergnault}, {Wallner}, {Waultier}, \&
  {Zanatta}}]{Auvergne2009}
{Auvergne}, M., {Bodin}, P., {Boisnard}, L., {et~al.} 2009, \aap, 506, 411

\bibitem[{{Balona}(2014)}]{Balona2014}
{Balona}, L.~A. 2014, \mnras, 437, 1476

\bibitem[{{Beck} {et~al.}(2014){Beck}, {Hambleton}, {Vos}, {Kallinger},
  {Bloemen}, {Tkachenko}, {Garc{\'{\i}}a}, {{\O}stensen}, {Aerts}, {Kurtz}, {De
  Ridder}, {Hekker}, {Pavlovski}, {Mathur}, {De Smedt}, {Derekas}, {Corsaro},
  {Mosser}, {Van Winckel}, {Huber}, {Degroote}, {Davies}, {Pr{\v s}a},
  {Debosscher}, {Elsworth}, {Nemeth}, {Siess}, {Schmid}, {P{\'a}pics}, {de
  Vries}, {van Marle}, {Marcos-Arenal}, \& {Lobel}}]{Beck2014}
{Beck}, P.~G., {Hambleton}, K., {Vos}, J., {et~al.} 2014, \aap, 564, A36

\bibitem[{{Beck} {et~al.}(2012){Beck}, {Montalban}, {Kallinger}, {De Ridder},
  {Aerts}, {Garc{\'{\i}}a}, {Hekker}, {Dupret}, {Mosser}, {Eggenberger},
  {Stello}, {Elsworth}, {Frandsen}, {Carrier}, {Hillen}, {Gruberbauer},
  {Christensen-Dalsgaard}, {Miglio}, {Valentini}, {Bedding}, {Kjeldsen},
  {Girouard}, {Hall}, \& {Ibrahim}}]{Beck2012}
{Beck}, P.~G., {Montalban}, J., {Kallinger}, T., {et~al.} 2012, \nat, 481, 55

\bibitem[{{Bedding} {et~al.}(2014){Bedding}, {Murphy}, {Colman}, \&
  {Kurtz}}]{Bedding2014}
{Bedding}, T.~R., {Murphy}, S.~J., {Colman}, I.~L., \& {Kurtz}, D.~W. 2014,
  ArXiv e-prints, arXiv:1411.1883

\bibitem[{{Borucki} {et~al.}(2010){Borucki}, {Koch}, {Basri}, {Batalha},
  {Brown}, {Caldwell}, {Caldwell}, {Christensen-Dalsgaard}, {Cochran},
  {DeVore}, {Dunham}, {Dupree}, {Gautier}, {Geary}, {Gilliland}, {Gould},
  {Howell}, {Jenkins}, {Kondo}, {Latham}, {Marcy}, {Meibom}, {Kjeldsen},
  {Lissauer}, {Monet}, {Morrison}, {Sasselov}, {Tarter}, {Boss}, {Brownlee},
  {Owen}, {Buzasi}, {Charbonneau}, {Doyle}, {Fortney}, {Ford}, {Holman},
  {Seager}, {Steffen}, {Welsh}, {Rowe}, {Anderson}, {Buchhave}, {Ciardi},
  {Walkowicz}, {Sherry}, {Horch}, {Isaacson}, {Everett}, {Fischer}, {Torres},
  {Johnson}, {Endl}, {MacQueen}, {Bryson}, {Dotson}, {Haas}, {Kolodziejczak},
  {Van Cleve}, {Chandrasekaran}, {Twicken}, {Quintana}, {Clarke}, {Allen},
  {Li}, {Wu}, {Tenenbaum}, {Verner}, {Bruhweiler}, {Barnes}, \&
  {Prsa}}]{Borucki2010}
{Borucki}, W.~J., {Koch}, D., {Basri}, G., {et~al.} 2010, Science, 327, 977

\bibitem[{{Bouabid} {et~al.}(2013){Bouabid}, {Dupret}, {Salmon},
  {Montalb{\'a}n}, {Miglio}, \& {Noels}}]{Bouabid2013}
{Bouabid}, M.-P., {Dupret}, M.-A., {Salmon}, S., {et~al.} 2013, \mnras, 429,
  2500

\bibitem[{{Bradley} {et~al.}(2015){Bradley}, {Guzik}, {Miles}, {Uytterhoeven},
  {Jackiewicz}, \& {Kinemuchi}}]{Bradley2015}
{Bradley}, P.~A., {Guzik}, J.~A., {Miles}, L.~F., {et~al.} 2015, \aj, 149, 68

\bibitem[{{Breger}(2000)}]{Breger2000a}
{Breger}, M. 2000, in Astronomical Society of the Pacific Conference Series,
  Vol. 210, Delta Scuti and Related Stars, ed. M.~{Breger} \& M.~{Montgomery},
  3

\bibitem[{{Breger} {et~al.}(1993){Breger}, {Stich}, {Garrido}, {Martin},
  {Jiang}, {Li}, {Hube}, {Ostermann}, {Paparo}, \& {Scheck}}]{Breger1993}
{Breger}, M., {Stich}, J., {Garrido}, R., {et~al.} 1993, \aap, 271, 482

\bibitem[{{Cantiello} {et~al.}(2014){Cantiello}, {Mankovich}, {Bildsten},
  {Christensen-Dalsgaard}, \& {Paxton}}]{Cantiello2014}
{Cantiello}, M., {Mankovich}, C., {Bildsten}, L., {Christensen-Dalsgaard}, J.,
  \& {Paxton}, B. 2014, \apj, 788, 93

\bibitem[{{Chaplin} \& {Miglio}(2013)}]{Chaplin2013}
{Chaplin}, W.~J. \& {Miglio}, A. 2013, \araa, 51, 353

\bibitem[{{Claret}(2003)}]{Claret2003}
{Claret}, A. 2003, \aap, 406, 623

\bibitem[{{Claret} \& {Bloemen}(2011)}]{Claret2011}
{Claret}, A. \& {Bloemen}, S. 2011, \aap, 529, A75

\bibitem[{{Debosscher} {et~al.}(2013){Debosscher}, {Aerts}, {Tkachenko},
  {Pavlovski}, {Maceroni}, {Kurtz}, {Beck}, {Bloemen}, {Degroote}, {Lombaert},
  \& {Southworth}}]{Debosscher2013}
{Debosscher}, J., {Aerts}, C., {Tkachenko}, A., {et~al.} 2013, \aap, 556, A56

\bibitem[{{Degroote} {et~al.}(2009){Degroote}, {Aerts}, {Ollivier}, {Miglio},
  {Debosscher}, {Cuypers}, {Briquet}, {Montalb{\'a}n}, {Thoul}, {Noels}, {De
  Cat}, {Balaguer-N{\'u}{\~n}ez}, {Maceroni}, {Ribas}, {Auvergne}, {Baglin},
  {Deleuil}, {Weiss}, {Jorda}, {Baudin}, \& {Samadi}}]{Degroote2009}
{Degroote}, P., {Aerts}, C., {Ollivier}, M., {et~al.} 2009, \aap, 506, 471

\bibitem[{{Deheuvels} {et~al.}(2015){Deheuvels}, {Ballot}, {Beck}, {Mosser},
  {{\O}stensen}, {Garc{\'{\i}}a}, \& {Goupil}}]{Deheuvels2015}
{Deheuvels}, S., {Ballot}, J., {Beck}, P.~G., {et~al.} 2015, \aap, 580, A96

\bibitem[{{Deheuvels} {et~al.}(2014){Deheuvels}, {Do{\u g}an}, {Goupil},
  {Appourchaux}, {Benomar}, {Bruntt}, {Campante}, {Casagrande}, {Ceillier},
  {Davies}, {De Cat}, {Fu}, {Garc{\'{\i}}a}, {Lobel}, {Mosser}, {Reese},
  {Regulo}, {Schou}, {Stahn}, {Thygesen}, {Yang}, {Chaplin},
  {Christensen-Dalsgaard}, {Eggenberger}, {Gizon}, {Mathis},
  {Molenda-{\.Z}akowicz}, \& {Pinsonneault}}]{Deheuvels2014}
{Deheuvels}, S., {Do{\u g}an}, G., {Goupil}, M.~J., {et~al.} 2014, \aap, 564,
  A27

\bibitem[{{Dupret} {et~al.}(2005){Dupret}, {Grigahc{\`e}ne}, {Garrido},
  {Gabriel}, \& {Scuflaire}}]{Dupret2005}
{Dupret}, M.-A., {Grigahc{\`e}ne}, A., {Garrido}, R., {Gabriel}, M., \&
  {Scuflaire}, R. 2005, \aap, 435, 927

\bibitem[{{Foreman-Mackey} {et~al.}(2013){Foreman-Mackey}, {Hogg}, {Lang}, \&
  {Goodman}}]{Foreman-Mackey2013}
{Foreman-Mackey}, D., {Hogg}, D.~W., {Lang}, D., \& {Goodman}, J. 2013, \pasp,
  125, 306

\bibitem[{Foreman-Mackey {et~al.}(2014)Foreman-Mackey, Price-Whelan, Ryan,
  Emily, Smith, Barbary, Hogg, \& Brewer}]{Foreman-Mackey2014}
Foreman-Mackey, D., Price-Whelan, A., Ryan, G., {et~al.} 2014, triangle.py
  v0.1.1

\bibitem[{{Fuller} {et~al.}(2015){Fuller}, {Cantiello}, {Lecoanet}, \&
  {Quataert}}]{Fuller2015}
{Fuller}, J., {Cantiello}, M., {Lecoanet}, D., \& {Quataert}, E. 2015, ArXiv
  e-prints, arXiv: 1502.07779

\bibitem[{{Fuller} \& {Lai}(2012)}]{Fuller2012}
{Fuller}, J. \& {Lai}, D. 2012, \mnras, 420, 3126

\bibitem[{{Goode} \& {Thompson}(1992)}]{Goode1992}
{Goode}, P.~R. \& {Thompson}, M.~J. 1992, \apj, 395, 307

\bibitem[{{Grevesse} {et~al.}(2007){Grevesse}, {Asplund}, \&
  {Sauval}}]{Grevesse2007}
{Grevesse}, N., {Asplund}, M., \& {Sauval}, A.~J. 2007, \ssr, 130, 105

\bibitem[{{Grigahc{\`e}ne} {et~al.}(2010){Grigahc{\`e}ne}, {Antoci}, {Balona},
  {Catanzaro}, {Daszy{\'n}ska-Daszkiewicz}, {Guzik}, {Handler}, {Houdek},
  {Kurtz}, {Marconi}, {Monteiro}, {Moya}, {Ripepi}, {Su{\'a}rez},
  {Uytterhoeven}, {Borucki}, {Brown}, {Christensen-Dalsgaard}, {Gilliland},
  {Jenkins}, {Kjeldsen}, {Koch}, {Bernabei}, {Bradley}, {Breger}, {Di
  Criscienzo}, {Dupret}, {Garc{\'{\i}}a}, {Garc{\'{\i}}a Hern{\'a}ndez},
  {Jackiewicz}, {Kaiser}, {Lehmann}, {Mart{\'{\i}}n-Ruiz}, {Mathias},
  {Molenda-{\.Z}akowicz}, {Nemec}, {Nuspl}, {Papar{\'o}}, {Roth}, {Szab{\'o}},
  {Suran}, \& {Ventura}}]{Grigahcene2010}
{Grigahc{\`e}ne}, A., {Antoci}, V., {Balona}, L., {et~al.} 2010, \apjl, 713,
  L192

\bibitem[{{Guzik} {et~al.}(2000){Guzik}, {Kaye}, {Bradley}, {Cox}, \&
  {Neuforge}}]{Guzik2000}
{Guzik}, J.~A., {Kaye}, A.~B., {Bradley}, P.~A., {Cox}, A.~N., \& {Neuforge},
  C. 2000, \apjl, 542, L57

\bibitem[{{Hadrava}(1995)}]{Hadrava1995}
{Hadrava}, P. 1995, \aaps, 114, 393

\bibitem[{{Hambleton} {et~al.}(2013){Hambleton}, {Kurtz}, {Pr{\v s}a}, {Guzik},
  {Pavlovski}, {Bloemen}, {Southworth}, {Conroy}, {Littlefair}, \&
  {Fuller}}]{Hambleton2013}
{Hambleton}, K.~M., {Kurtz}, D.~W., {Pr{\v s}a}, A., {et~al.} 2013, \mnras,
  434, 925

\bibitem[{{Huber} {et~al.}(2014){Huber}, {Silva Aguirre}, {Matthews},
  {Pinsonneault}, {Gaidos}, {Garc{\'{\i}}a}, {Hekker}, {Mathur}, {Mosser},
  {Torres}, {Bastien}, {Basu}, {Bedding}, {Chaplin}, {Demory}, {Fleming},
  {Guo}, {Mann}, {Rowe}, {Serenelli}, {Smith}, \& {Stello}}]{Huber2014}
{Huber}, D., {Silva Aguirre}, V., {Matthews}, J.~M., {et~al.} 2014, \apjs, 211,
  2

\bibitem[{{Ilijic} {et~al.}(2004){Ilijic}, {Hensberge}, {Pavlovski}, \&
  {Freyhammer}}]{Ilijic2004}
{Ilijic}, S., {Hensberge}, H., {Pavlovski}, K., \& {Freyhammer}, L.~M. 2004, in
  Astronomical Society of the Pacific Conference Series, Vol. 318,
  Spectroscopically and Spatially Resolving the Components of the Close Binary
  Stars, ed. R.~W. {Hilditch}, H.~{Hensberge}, \& K.~{Pavlovski}, 111--113

\bibitem[{{Kaye} {et~al.}(1999){Kaye}, {Handler}, {Krisciunas}, {Poretti}, \&
  {Zerbi}}]{Kaye1999}
{Kaye}, A.~B., {Handler}, G., {Krisciunas}, K., {Poretti}, E., \& {Zerbi},
  F.~M. 1999, \pasp, 111, 840

\bibitem[{{Keen} {et~al.}(2015){Keen}, {Bedding}, {Murphy}, {Schmid}, {Aerts},
  {Tkachenko}, {Ouazzani}, \& {Kurtz}}]{Keen2015}
{Keen}, M.~A., {Bedding}, T.~R., {Murphy}, S.~J., {et~al.} 2015, ArXiv
  e-prints, arXiv:1509.03317

\bibitem[{{Kurtz} {et~al.}(2014){Kurtz}, {Saio}, {Takata}, {Shibahashi},
  {Murphy}, \& {Sekii}}]{Kurtz2014}
{Kurtz}, D.~W., {Saio}, H., {Takata}, M., {et~al.} 2014, \mnras, 444, 102

\bibitem[{{Kurtz} {et~al.}(2015){Kurtz}, {Shibahashi}, {Murphy}, {Bedding}, \&
  {Bowman}}]{Kurtz2015}
{Kurtz}, D.~W., {Shibahashi}, H., {Murphy}, S.~J., {Bedding}, T.~R., \&
  {Bowman}, D.~M. 2015, \mnras, 450, 3015

\bibitem[{{Ledoux}(1951)}]{Ledoux1951}
{Ledoux}, P. 1951, \apj, 114, 373

\bibitem[{{Lehmann} {et~al.}(2011){Lehmann}, {Tkachenko}, {Semaan},
  {Guti{\'e}rrez-Soto}, {Smalley}, {Briquet}, {Shulyak}, {Tsymbal}, \& {De
  Cat}}]{Lehmann2011}
{Lehmann}, H., {Tkachenko}, A., {Semaan}, T., {et~al.} 2011, \aap, 526, A124

\bibitem[{{Loumos} \& {Deeming}(1978)}]{Loumos1978}
{Loumos}, G.~L. \& {Deeming}, T.~J. 1978, \apss, 56, 285

\bibitem[{{Lucy}(1967)}]{Lucy1967}
{Lucy}, L.~B. 1967, \zap, 65, 89

\bibitem[{{Maceroni} {et~al.}(2014){Maceroni}, {Lehmann}, {da Silva},
  {Montalb{\'a}n}, {Lee}, {Ak}, {Deshpande}, {Yakut}, {Debosscher}, {Guo},
  {Kim}, {Lee}, \& {Southworth}}]{Maceroni2014}
{Maceroni}, C., {Lehmann}, H., {da Silva}, R., {et~al.} 2014, \aap, 563, A59

\bibitem[{{Maceroni} {et~al.}(2013){Maceroni}, {Montalb{\'a}n}, {Gandolfi},
  {Pavlovski}, \& {Rainer}}]{Maceroni2013}
{Maceroni}, C., {Montalb{\'a}n}, J., {Gandolfi}, D., {Pavlovski}, K., \&
  {Rainer}, M. 2013, \aap, 552, A60

\bibitem[{{Maceroni} {et~al.}(2009){Maceroni}, {Montalb{\'a}n}, {Michel},
  {Harmanec}, {Prsa}, {Briquet}, {Niemczura}, {Morel}, {Ladjal}, {Auvergne},
  {Baglin}, {Baudin}, {Catala}, {Samadi}, \& {Aerts}}]{Maceroni2009}
{Maceroni}, C., {Montalb{\'a}n}, J., {Michel}, E., {et~al.} 2009, \aap, 508,
  1375

\bibitem[{{Miglio} {et~al.}(2008){Miglio}, {Montalb{\'a}n}, {Noels}, \&
  {Eggenberger}}]{Miglio2008}
{Miglio}, A., {Montalb{\'a}n}, J., {Noels}, A., \& {Eggenberger}, P. 2008,
  \mnras, 386, 1487

\bibitem[{{Montgomery} \& {O'Donoghue}(1999)}]{Montgomery1999}
{Montgomery}, M.~H. \& {O'Donoghue}, D. 1999, Delta Scuti Star Newsletter, 13,
  28

\bibitem[{{Moravveji} {et~al.}(2015){Moravveji}, {Aerts}, {Papics}, {Andres
  Triana}, \& {Vandoren}}]{Moravveji2015}
{Moravveji}, E., {Aerts}, C., {Papics}, P.~I., {Andres Triana}, S., \&
  {Vandoren}, B. 2015, \aap, 580, A27

\bibitem[{{Mosser} {et~al.}(2012){Mosser}, {Goupil}, {Belkacem}, {Marques},
  {Beck}, {Bloemen}, {De Ridder}, {Barban}, {Deheuvels}, {Elsworth}, {Hekker},
  {Kallinger}, {Ouazzani}, {Pinsonneault}, {Samadi}, {Stello}, {Garc{\'{\i}}a},
  {Klaus}, {Li}, {Mathur}, \& {Morris}}]{Mosser2012}
{Mosser}, B., {Goupil}, M.~J., {Belkacem}, K., {et~al.} 2012, \aap, 548, A10

\bibitem[{{Murphy} {et~al.}(2014){Murphy}, {Bedding}, {Shibahashi}, {Kurtz}, \&
  {Kjeldsen}}]{Murphy2014}
{Murphy}, S.~J., {Bedding}, T.~R., {Shibahashi}, H., {Kurtz}, D.~W., \&
  {Kjeldsen}, H. 2014, \mnras, 441, 2515

\bibitem[{{Murphy} \& {Shibahashi}(2015)}]{Murphy2015}
{Murphy}, S.~J. \& {Shibahashi}, H. 2015, \mnras, 450, 4475

\bibitem[{{P{\'a}pics}(2012)}]{Papics2012a}
{P{\'a}pics}, P.~I. 2012, Astronomische Nachrichten, 333, 1053

\bibitem[{{P{\'a}pics} {et~al.}(2013){P{\'a}pics}, {Tkachenko}, {Aerts},
  {Briquet}, {Marcos-Arenal}, {Beck}, {Uytterhoeven}, {Trivi{\~n}o Hage},
  {Southworth}, {Clubb}, {Bloemen}, {Degroote}, {Jackiewicz}, {McKeever}, {Van
  Winckel}, {Niemczura}, {Gameiro}, \& {Debosscher}}]{Papics2013}
{P{\'a}pics}, P.~I., {Tkachenko}, A., {Aerts}, C., {et~al.} 2013, \aap, 553,
  A127

\bibitem[{{P{\'a}pics} {et~al.}(2015){P{\'a}pics}, {Tkachenko}, {Aerts}, {Van
  Reeth}, {De Smedt}, {Hillen}, {{\O}stensen}, \& {Moravveji}}]{Papics2015}
{P{\'a}pics}, P.~I., {Tkachenko}, A., {Aerts}, C., {et~al.} 2015, \apjl, 803,
  L25

\bibitem[{{Pr{\v s}a} \& {Zwitter}(2005)}]{Prsa2005}
{Pr{\v s}a}, A. \& {Zwitter}, T. 2005, \apj, 628, 426

\bibitem[{{Raskin} {et~al.}(2011){Raskin}, {van Winckel}, {Hensberge},
  {Jorissen}, {Lehmann}, {Waelkens}, {Avila}, {de Cuyper}, {Degroote},
  {Dubosson}, {Dumortier}, {Fr{\'e}mat}, {Laux}, {Michaud}, {Morren}, {Perez
  Padilla}, {Pessemier}, {Prins}, {Smolders}, {van Eck}, \&
  {Winkler}}]{Raskin2011}
{Raskin}, G., {van Winckel}, H., {Hensberge}, H., {et~al.} 2011, \aap, 526, A69

\bibitem[{{Rucinski}(1989)}]{Rucinski1989}
{Rucinski}, S.~M. 1989, Comments on Astrophysics, 14, 79

\bibitem[{{Saio}(1981)}]{Saio1981}
{Saio}, H. 1981, \apj, 244, 299

\bibitem[{{Saio} {et~al.}(2015){Saio}, {Kurtz}, {Takata}, {Shibahashi},
  {Murphy}, {Sekii}, \& {Bedding}}]{Saio2015}
{Saio}, H., {Kurtz}, D.~W., {Takata}, M., {et~al.} 2015, \mnras, 447, 3264

\bibitem[{{Schwarzenberg-Czerny}(2003)}]{Schwarzenberg-Czerny2003}
{Schwarzenberg-Czerny}, A. 2003, in Astronomical Society of the Pacific
  Conference Series, Vol. 292, Interplay of Periodic, Cyclic and Stochastic
  Variability in Selected Areas of the H-R Diagram, ed. C.~{Sterken}, 383

\bibitem[{{Shibahashi} \& {Kurtz}(2012)}]{Shibahashi2012}
{Shibahashi}, H. \& {Kurtz}, D.~W. 2012, \mnras, 422, 738

\bibitem[{{Shibahashi} {et~al.}(2015){Shibahashi}, {Kurtz}, \&
  {Murphy}}]{Shibahashi2015}
{Shibahashi}, H., {Kurtz}, D.~W., \& {Murphy}, S.~J. 2015, \mnras, 450, 3999

\bibitem[{{Simon} \& {Sturm}(1994)}]{Simon1994}
{Simon}, K.~P. \& {Sturm}, E. 1994, \aap, 281, 286

\bibitem[{{Southworth}(2012)}]{Southworth2012}
{Southworth}, J. 2012, in Orbital Couples: Pas de Deux in the Solar System and
  the Milky Way, ed. F.~{Arenou} \& D.~{Hestroffer}, 51--58

\bibitem[{{Tassoul}(1980)}]{Tassoul1980}
{Tassoul}, M. 1980, \apjs, 43, 469

\bibitem[{{Telting} {et~al.}(2014){Telting}, {Baran}, {Nemeth}, {{\O}stensen},
  {Kupfer}, {Macfarlane}, {Heber}, {Aerts}, \& {Geier}}]{Telting2014}
{Telting}, J.~H., {Baran}, A.~S., {Nemeth}, P., {et~al.} 2014, \aap, 570, A129

\bibitem[{{Tkachenko}(2015)}]{Tkachenko2015}
{Tkachenko}, A. 2015, ArXiv e-prints, arXiv:1507.02864

\bibitem[{{Tkachenko} {et~al.}(2013){Tkachenko}, {Aerts}, {Yakushechkin},
  {Debosscher}, {Degroote}, {Bloemen}, {P{\'a}pics}, {de Vries}, {Lombaert},
  {Hrudkova}, {Fr{\'e}mat}, {Raskin}, \& {Van Winckel}}]{Tkachenko2013a}
{Tkachenko}, A., {Aerts}, C., {Yakushechkin}, A., {et~al.} 2013, \aap, 556, A52

\bibitem[{{Triana} {et~al.}(2015){Triana}, {Moravveji}, {P{\'a}pics}, {Aerts},
  {Kawaler}, \& {Christensen-Dalsgaard}}]{Triana2015}
{Triana}, S.~A., {Moravveji}, E., {P{\'a}pics}, P.~I., {et~al.} 2015, ArXiv
  e-prints, arXiv:1507.04574

\bibitem[{{Uytterhoeven} {et~al.}(2011){Uytterhoeven}, {Moya},
  {Grigahc{\`e}ne}, {Guzik}, {Guti{\'e}rrez-Soto}, {Smalley}, {Handler},
  {Balona}, {Niemczura}, {Fox Machado}, {Benatti}, {Chapellier}, {Tkachenko},
  {Szab{\'o}}, {Su{\'a}rez}, {Ripepi}, {Pascual}, {Mathias},
  {Mart{\'{\i}}n-Ru{\'{\i}}z}, {Lehmann}, {Jackiewicz}, {Hekker},
  {Gruberbauer}, {Garc{\'{\i}}a}, {Dumusque}, {D{\'{\i}}az-Fraile}, {Bradley},
  {Antoci}, {Roth}, {Leroy}, {Murphy}, {De Cat}, {Cuypers}, {Kjeldsen},
  {Christensen-Dalsgaard}, {Breger}, {Pigulski}, {Kiss}, {Still}, {Thompson},
  \& {van Cleve}}]{Uytterhoeven2011}
{Uytterhoeven}, K., {Moya}, A., {Grigahc{\`e}ne}, A., {et~al.} 2011, \aap, 534,
  A125

\bibitem[{{Van Reeth} {et~al.}(2015{\natexlab{a}}){Van Reeth}, {Tkachenko},
  {Aerts}, {P{\'a}pics}, {Degroote}, {Debosscher}, {Zwintz}, {Bloemen}, {De
  Smedt}, {Hrudkova}, {Raskin}, \& {Van Winckel}}]{VanReeth2015a}
{Van Reeth}, T., {Tkachenko}, A., {Aerts}, C., {et~al.} 2015{\natexlab{a}},
  \aap, 574, A17

\bibitem[{{Van Reeth} {et~al.}(2015{\natexlab{b}}){Van Reeth}, {Tkachenko},
  {Aerts}, {P{\'a}pics}, {Triana}, {Zwintz}, {Degroote}, {Debosscher},
  {Bloemen}, {Schmid}, {De Smedt}, {Fremat}, {Fuentes}, {Homan}, {Hrudkova},
  {Karjalainen}, {Lombaert}, {Nemeth}, {{\O}stensen}, {Van De Steene}, {Vos},
  {Raskin}, \& {Van Winckel}}]{VanReeth2015b}
{Van Reeth}, T., {Tkachenko}, A., {Aerts}, C., {et~al.} 2015{\natexlab{b}},
  \apjs, 218, 27

\bibitem[{{Walker} {et~al.}(2003){Walker}, {Matthews}, {Kuschnig}, {Johnson},
  {Rucinski}, {Pazder}, {Burley}, {Walker}, {Skaret}, {Zee}, {Grocott},
  {Carroll}, {Sinclair}, {Sturgeon}, \& {Harron}}]{Walker2003}
{Walker}, G., {Matthews}, J., {Kuschnig}, R., {et~al.} 2003, \pasp, 115, 1023

\bibitem[{{Welsh} {et~al.}(2011){Welsh}, {Orosz}, {Aerts}, {Brown},
  {Brugamyer}, {Cochran}, {Gilliland}, {Guzik}, {Kurtz}, {Latham}, {Marcy},
  {Quinn}, {Zima}, {Allen}, {Batalha}, {Bryson}, {Buchhave}, {Caldwell},
  {Gautier}, {Howell}, {Kinemuchi}, {Ibrahim}, {Isaacson}, {Jenkins}, {Prsa},
  {Still}, {Street}, {Wohler}, {Koch}, \& {Borucki}}]{Welsh2011}
{Welsh}, W.~F., {Orosz}, J.~A., {Aerts}, C., {et~al.} 2011, \apjs, 197, 4

\bibitem[{{Wilson}(1979)}]{Wilson1979}
{Wilson}, R.~E. 1979, \apj, 234, 1054

\bibitem[{{Wilson}(1990)}]{Wilson1990}
{Wilson}, R.~E. 1990, \apj, 356, 613

\bibitem[{{Wilson} \& {Devinney}(1971)}]{Wilson1971}
{Wilson}, R.~E. \& {Devinney}, E.~J. 1971, \apj, 166, 605

\bibitem[{{Zwintz} {et~al.}(2014){Zwintz}, {Fossati}, {Ryabchikova},
  {Guenther}, {Aerts}, {Barnes}, {Theme{\ss}l}, {Lorenz}, {Cameron},
  {Kuschnig}, {Pollack-Drs}, {Moravveji}, {Baglin}, {Matthews}, {Moffat},
  {Poretti}, {Rainer}, {Rucinski}, {Sasselov}, \& {Weiss}}]{Zwintz2014}
{Zwintz}, K., {Fossati}, L., {Ryabchikova}, T., {et~al.} 2014, Science, 345,
  550

\end{thebibliography}

\appendix
\section{Frequency analysis results}

\begin{longtable}{r r r r r r r r r r l}
\caption{\label{tab:fourier_results}Results of the pulsation frequency analysis for KIC\,10080943. Values listed are: Frequency $f$, amplitude $A$, and phase $\phi$, their respective uncertainties $\epsilon_f$, $\epsilon_A$, and $\epsilon_\phi$, as well as the $S/N$ calculated in a frequency range of 1~d$^{-1}$, 3~d$^{-1}$, and the complete range up to the Nyquist frequency. Period spacing doublets (PS d) and triplets (PS t), frequency splittings (FS) and combination frequencies are identified by comments.}\\
\hline\hline
 & \multicolumn{1}{c}{$f$} & \multicolumn{1}{c}{$\epsilon_f$} & \multicolumn{1}{c}{$A$} & \multicolumn{1}{c}{$\epsilon_A$} & \multicolumn{1}{c}{$\phi$} & \multicolumn{1}{c}{$\epsilon_\phi$} & \multicolumn{3}{c}{$S/N$} & Comment \\
 & \multicolumn{1}{c}{d$^{-1}$} & \multicolumn{1}{c}{d$^{-1}$} & \multicolumn{1}{c}{ppm} & \multicolumn{1}{c}{ppm} & \multicolumn{1}{c}{$2\pi$/rad} & \multicolumn{1}{c}{$2\pi$/rad} & \multicolumn{1}{c}{1~d$^{-1}$} & \multicolumn{1}{c}{3~d$^{-1}$} & \multicolumn{1}{c}{Full} & \\
 \hline
\endfirsthead
\caption{continued.}\\
\hline\hline
 & \multicolumn{1}{c}{$f$} & \multicolumn{1}{c}{$\epsilon_f$} & \multicolumn{1}{c}{$A$} & \multicolumn{1}{c}{$\epsilon_A$} & \multicolumn{1}{c}{$\phi$} & \multicolumn{1}{c}{$\epsilon_\phi$} & \multicolumn{3}{c}{$S/N$} & Comment \\
 & \multicolumn{1}{c}{d$^{-1}$} & \multicolumn{1}{c}{d$^{-1}$} & \multicolumn{1}{c}{ppm} & \multicolumn{1}{c}{ppm} & \multicolumn{1}{c}{$2\pi$/rad} & \multicolumn{1}{c}{$2\pi$/rad} & \multicolumn{1}{c}{1~d$^{-1}$} & \multicolumn{1}{c}{3~d$^{-1}$} & \multicolumn{1}{c}{Full} & \\
\hline
\endhead
\hline
\endfoot
$f_\mathrm{1}$ & 1.059380 & 0.000009 & 1977.8 & 47.3 & $-0.060$ & 0.024 & 220.0 & 244.1 & 552.9 & PS d; $m=-1$ \\
$f_\mathrm{2}$ & 1.150808 & 0.000009 & 1772.4 & 44.4 & $0.486$ & 0.025 & 204.9 & 218.7 & 495.5 & PS d; $m=1$ \\
$f_\mathrm{3}$ & 0.88692 & 0.00001 & 1532.7 & 42.0 & $-0.198$ & 0.027 & 158.3 & 189.1 & 428.5 & PS d; $m=-1$ \\
$f_\mathrm{4}$ & 13.947586 & 0.000002 & 1360.9 & 7.7 & $0.140$ & 0.006 & 861.7 & 788.8 & 380.5 &  \\
$f_\mathrm{5}$ & 1.07358 & 0.00001 & 1358.9 & 40.5 & $-0.244$ & 0.030 & 152.4 & 167.7 & 379.9 & PS d; $m=1$ \\
$f_\mathrm{6}$ & 0.948204 & 0.000008 & 1340.3 & 29.4 & $-0.477$ & 0.022 & 142.5 & 165.4 & 374.7 & PS d; $m=-1$ \\
$f_\mathrm{7}$ & 1.23814 & 0.00001 & 1324.2 & 37.1 & $-0.459$ & 0.028 & 157.9 & 163.4 & 370.2 & PS d; $m=1$ \\
$f_\mathrm{8}$ & 3.33350 & 0.00001 & 1321.8 & 38.6 & $0.400$ & 0.029 & 213.9 & 227.2 & 369.5 &  \\
\hline
\end{longtable}

\end{document}